\documentclass[a4paper,11pt]{article}
\pdfoutput=1

\usepackage{jheppub}

\usepackage{amsmath}
\input epsf
\usepackage{epsfig}
\usepackage{amssymb}
\usepackage{graphics}
\usepackage{adjustbox}
\usepackage{tikz}
\usepackage{comment}
\usepackage{mathtools}
\usepackage[active]{srcltx}

\usepackage[T1]{fontenc} 
\usepackage{array}

\usepackage{slashed}

\usepackage{shuffle}

\usetikzlibrary{decorations.markings}

\newcommand{\be}{\begin{equation}}
\newcommand{\ee}{\end{equation}}
\newcommand{\bea}{\begin{eqnarray}}
\newcommand{\eea}{\end{eqnarray}}
\newcommand{\bean}{\begin{eqnarray*}}
\newcommand{\eean}{\end{eqnarray*}}
\newcommand{\nn}{\nonumber \\}

\def\O #1{\overline{#1}}

\def\W #1{\widetilde{#1}}

\def\eref#1{(\ref{#1})}

\def\a{{\alpha}}

\def\eps{\epsilon}

\def\YM{{\tiny\mbox{YM}}}

\def\EYM{{\tiny \mbox{EYM}}}

\def\YM{{\tiny\mbox{YM}}}

\def\Label#1{\label{#1}}%

\newcommand {\Pf}  {\text{Pf}\,}
\newcommand {\Pfp}  {\text{Pf}\,'}
\newcommand {\PT} {\text{PT}\,}

\title{Expanding Einstein-Yang-Mills  by Yang-Mills in CHY frame}

\author[a]{Fei Teng}
\emailAdd{Fei.Teng@utah.edu}
\author[b,c]{and Bo Feng}
\emailAdd{fengbo@zju.edu.cn}

\affiliation[a]{Department of Physics and Astronomy, University of Utah,\\ 115 South 1400 East, Salt Lake City, UT 84112, USA}
\affiliation[b]{Zhejiang Institute of Modern Physics, Department of Physics, Zhejiang University, \\
No. 38 Zheda Road, Hangzhou 310027, P.R. China}
\affiliation[c]{Center of Mathematical Science, Zhejiang University,\\
No. 38 Zheda Road, Hangzhou 310027, P.R. China.}

\abstract{Using the Cachazo-He-Yuan (CHY) formalism, we prove a recursive expansion of tree level single trace Einstein-Yang-Mills (EYM) amplitudes with an arbitrary number of gluons and gravitons, which is valid for general spacetime dimensions and any helicity configurations. The recursion is written in terms of fewer-graviton EYM amplitudes and pure Yang-Mills (YM) amplitudes, which can be further carried out until we reach an expansion in terms of pure YM amplitudes in Kleiss-Kuijf (KK) basis. Our expansion then generates naturally a spanning tree structure rooted on gluons whose vertices are gravitons. We further propose a set of graph theoretical rules based on spanning trees that evaluate directly the pure YM expansion coefficients.}


\begin{document}
\maketitle
\flushbottom

\section{Introduction}

Recently, non-trivial relations between Einstein-Yang-Mills (EYM) and pure Yang-Mills (YM) amplitudes\footnote{Unless otherwise noted, we only consider tree level single trace color ordered EYM amplitudes.} have been studied in several papers~\cite{Stieberger:2016lng,Nandan:2016pya,delaCruz:2016gnm,Schlotterer:2016cxa,Du:2016wkt,Nandan:2016ohb,He:2016mzd,Fu:2017uzt}. These works have more or less the same goal: to find the relations that expand the EYM amplitude with $n$ gluons and $m$ gravitons in terms of a linear sum of $(n+m)$-point pure YM amplitudes, although it is studied from different points of view. The starting point of our paper is the Cachazo-He-Yuan (CHY) formulation~\cite{Cachazo:2013iaa,Cachazo:2013gna,Cachazo:2013hca,Cachazo:2013iea,Cachazo:2014nsa,Cachazo:2014xea}, which is the same as most of the above mentioned discussions.
The key idea in this pursuit is to use various identities to reformulate the CHY integrand for EYM into an appropriate form. In particular, the cross-ratio identity and other off-shell identities of integrands~\cite{Bjerrum-Bohr:2016juj,Cardona:2016gon} play a very crucial rule\footnote{See Appendix of~\cite{Huang:2017ydz} for summary of these identities.}. With the help of these techniques, the explicit expansions of EYM amplitudes with up to three gravitons have been provided in~\cite{Nandan:2016pya}. Based on these explicit results, an recursive expansion for EYM amplitudes with an arbitrary number of gravitons has been conjectured in~\cite{Fu:2017uzt}, using the gauge invariance principle advocated in~\cite{Boels:2016xhc,Arkani-Hamed:2016rak,Rodina:2016mbk,Rodina:2016jyz}.

One of the main subjects in this paper is to investigate directly in the CHY frame the aforementioned recursive structure proposed in~\cite{Fu:2017uzt}. Especially, we want to see how this structure appears at the level of CHY integrands. Intuitively, we believe that the CHY formalism sets one of the best frameworks to understand such a recursive structure. The reason is that the information related to graviton polarizations is packed into a Pfaffian, which can be recursively evaluated by the Laplace expansion: for a $2n\times 2n$ antisymmetric matrix $M=(m_{ij})$, its Pfaffian can be calculated by
\begin{equation}
\label{eq:laplace}
\Pf(M)=\sum_{j=1,\,j\neq i}^{2n}(-1)^{i+j+1+\theta(i-j)}m_{ij}\,\Pf(M_{ij}^{ij})\,,
\end{equation}
where $\theta(i-j)$ is a step function and $M_{ij}^{ij}$ denotes the matrix obtained from $M$ by deleting the $i$-th and $j$-th row and column. As we will see later in Section~\ref{sec:example} and \ref{sec:recursive}, such a Laplace expansion of the $m$-graviton EYM integrand always contains an $(m-1)$-graviton integrand, plus other algebraic minors. Then if we can prove that those additional algebraic minors reduce to a linear combination of pure YM and EYM integrands with $m-2$ gravitons at most, we essentially derive a recursive structure! The proof requires an elegant arrangement of the Laplace expansion for Pfaffian, and a proper use of various on-shell and off-shell identities, to be spelled out later. This technique can possibly be generalized to a variety of theories constructed in, for example,~\cite{Cachazo:2014nsa,Cachazo:2014xea,Cachazo:2016njl}.

Knowing the recursive relation, we can perform the expansion of generic EYM amplitudes level by level in a well-controlled way. Eventually, this process leads to an expansion in terms of pure YM Kleiss-Kuijf (KK) basis~\cite{Kleiss:1988ne}. Although the final result involves a huge number of terms, it is actually very well-organized and enjoys a very elegant spanning tree structure. We find that our recursive relation derives an algorithm that evaluates directly the spanning trees, yielding the expansion coefficients we want. Spanning tree structures have appeared in the literature before for EYM and gravity in four dimensional spacetime with the MHV helicity configuration~\cite{Nguyen:2009jk,Feng:2012sy,Du:2016wkt}. In contrary, the algorithm we propose here works in arbitrary spacetime dimensions for EYM amplitudes with general helicity configurations. {Since EYM amplitudes can be realized as a double copy of YM and YM-scalar, even at multitrace levels~\cite{Chiodaroli:2014xia,Chiodaroli:2015rdg,Cachazo:2014xea}, a very promising application of this technique will be the direct evaluation of Bern-Carrasco-Johansson (BCJ) numerators~\cite{Bern:2008qj,Bern:2010ue} for YM and YM-scalar.}

The structure of this paper is the following. In Section~\ref{sec:integrand}, we give the essential background knowledge of the CHY formalism. The emphasis is put on the definition of the tree level YM and single trace EYM integrand. Then in Section~\ref{sec:example} we give several explicit examples, up to four gravitons, on how to expand the EYM integrands, such that one can clearly observe the pattern and summarize it into a recursive relation for generic cases, which is then proved by induction in Section~\ref{sec:recursive}. We propose in Section~\ref{sec:GraphicRules} a set of graphic rules based on spanning trees to evaluate directly the expansion coefficients of EYM amplitudes in terms of pure YM ones in the KK basis~\cite{Kleiss:1988ne}. These rules of course come from the recursive relation we proved in Section~\ref{sec:recursive}, which is demonstrated in Appendix~\ref{sec:derivation}. Finally, we give the conclusion and discussion in Section~\ref{sec:conclusion}.

\section{CHY integrand for Yang-Mills, Einstein-Yang-Mills and gravity}\label{sec:integrand}
The central object we are going to study is the tree-level single trace color ordered EYM amplitude with $n$ gluons and $m$ gravitons, denoted by 
\begin{equation}
    A_{n,m}^{\text{EYM}}(12\ldots n\,|\,h_{m}h_{m-1}\ldots h_{1})\,.
\end{equation}
For convenience, we define the gluon and graviton set as $$\mathsf{G}=\{12\ldots n\}\qquad\mathsf{H}=\{h_{m}h_{m-1}\ldots h_{1}\}\,.$$
The CHY integrand for this amplitude is \cite{Cachazo:2014nsa}
\begin{equation}
    \mathcal{I}_{\text{EYM}}(12\ldots n\,|\,h_{m}h_{m-1}\ldots h_{1})=\PT(12\ldots n)\,\Pf(\Psi_{\mathsf{H}})\,\Pfp(\Psi)\,,\Label{EYM-CHY}
\end{equation}
where the Parke-Taylor factor:
\begin{equation}
    \PT(12\ldots n)=\frac{1}{\sigma_{12}\sigma_{23}\ldots\sigma_{n1}}\qquad\sigma_{ij}\equiv\sigma_{i}-\sigma_{j}
\end{equation}
encodes the color ordering of the gluons. The $2n\times 2n$ matrix $\Psi$ is defined as:
\begin{equation}
    \Psi=\left(\begin{array}{cc}
        A & -C^{T} \\
        C & B \\
    \end{array}\right)\,.
\end{equation}
The three $n\times n$ matrices $A$, $B$ and $C$ have the form:
\begin{align}
& A_{ab}=\left\{\begin{array}{>{\displaystyle}c @{\hspace{1em}} >{\displaystyle}l}
\frac{k_a\cdot k_b}{\sigma_{ab}} & a\neq b\\
0 & a=b \\
\end{array}\right.&
& B_{ab}=\left\{\begin{array}{>{\displaystyle}c @{\hspace{1em}} >{\displaystyle}l}
\frac{\epsilon_{a}\cdot\epsilon_{b}}{\sigma_{ab}} & a\neq b\\
0 & a=b \\
\end{array}\right.&
& C_{ab}=\left\{\begin{array}{>{\displaystyle}l @{\hspace{1em}} >{\displaystyle}l}
\frac{\epsilon_{a}\cdot k_{b}}{\sigma_{ab}} & a\neq b\\
-\sum_{c\neq a}\frac{\epsilon_{a}\cdot k_{c}}{\sigma_{ac}} & a=b \\
\end{array}\right.\,,
\label{eq:ABC}
\end{align}
where the indices $a$, $b$ and $c$ takes value within $\mathsf{G}\cup\mathsf{H}$. The reduced Pfaffian of $\Psi$ is defined as:
\begin{equation}
    \Pfp(\Psi)=\frac{(-1)^{i+j}}{\sigma_{ij}}\Pf(\Psi_{ij}^{ij})\qquad 1\leqslant i<j\leqslant n+m\,.
\end{equation}
The matrix $\Psi_{ij}^{ij}$ is reduced from $\Psi$ by deleting the $i$-th and $j$-th  row and column. Finally, the $2m\times 2m$ matrix $\Psi_{\mathsf{H}}$ is defined as:
\begin{equation}
    \Psi_{\mathsf{H}}=\left(\begin{array}{cc}
        A_{\mathsf{H}} & -(C_{\mathsf{H}})^{T} \\
        C_{\mathsf{H}} & B_{\mathsf{H}} \\
    \end{array}\right)\,,
\end{equation}
where $A_{\mathsf{H}}$, $B_{\mathsf{H}}$ and $C_{\mathsf{H}}$ are respectively the $m\times m$ submatrices of $A$, $B$ and $C$ whose rows and columns take value only within $\mathsf{H}$. In order to obtain the amplitude $A^{\rm EYM}_{n,m}$, we integrate over the measure $\Omega_{\text{CHY}}$ that localizes all the $\sigma$'s to the solutions of the scattering equation for $n+m$ massless particles:
\begin{equation}
\label{eq:CHYintegration}
    A_{n,m}^{\text{EYM}}(12\ldots n\,|\,h_mh_{m-1}\ldots h_{1})=(-1)^{\frac{m(m+1)}{2}}\int{\Omega_{\text{CHY}}}\,\mathcal{I}_{\text{EYM}}(12\ldots n\,|\,h_{m}h_{m-1}\ldots h_{1})\,.
\end{equation}
The expression of $\Omega_{\text{CHY}}$ is not needed in this work, and we refer the interested readers to the original works of CHY~\cite{Cachazo:2013iaa,Cachazo:2013gna,Cachazo:2013hca,Cachazo:2013iea,Cachazo:2014nsa,Cachazo:2014xea}. Similarly, we have the following integrands for tree-level YM and Einstein gravity:
\begin{equation}
    \mathcal{I}_{\text{YM}}(12\ldots n)=\PT(12\ldots n)\,\Pfp(\Psi)\qquad\mathcal{I}_{\text{GR}}(12\ldots n)=\Pfp(\Psi)\times\Pfp(\Psi)\,,\Label{YM-CHY}
\end{equation}
from which the amplitudes are obtained through a similar integration as in Eq.~\eqref{eq:CHYintegration}. Then Eq.~\eqref{EYM-CHY} and \eqref{YM-CHY} suggest that there should exist an expansion of the tree-level EYM integrands in terms of the YM ones, namely:
%
\be \PT(12\ldots n)\,\Pf(\Psi_{\mathsf{H}})=\sum_{\a\in S_{n+m-2}}C_\a\,\PT(1,\a\{2...n-1,h_1...h_m\}, n)\,,\Label{CHY-exp}\ee
where the equal sign holds when the $\sigma_i$'s are solutions to the scattering equations.

The general expansion\footnote{The results with one, two and three gravitons were first worked out in~\cite{Nandan:2016pya},
but the forms cited here were given in~\cite{Fu:2017uzt}.} at the amplitude level has been proposed in a recent paper~\cite{Fu:2017uzt} based on the gauge invariance. Let us first recall some examples:
\begin{itemize}

\item With only one graviton, the expansion is given by:
\bea A^{\EYM}_{n,1}(1,2,\ldots,n\,|\,p)=\sum_{\shuffle} (
\eps_p \cdot Y_p)
A_{n+1}^{\YM}(1,\{2,\ldots,n-1\}\shuffle\{p\},n)\,,\Label{EYM1gra}\eea
{where $\sum_\shuffle$ is over the shuffle product $\pmb{\rho}\shuffle\pmb{\omega}$ of two ordered sets $\pmb{\rho}$ and $\pmb{\omega}$ (i.e.,  all the permutations of $\pmb{\rho}\cup\pmb{\omega}$ that preserving the ordering of $\pmb{\rho}$ and $\pmb{\omega}$ respectively),} and $Y_p$ denotes the sum of the momenta of all the gluons ahead of the leg $p$ in the color-ordered YM amplitude. Thus $Y_p$ depends on the orderings contained in the shuffle product. We keep this dependence implicit throughout this work in order not to make our notation too cluttered.

\item With two gravitons, the expansion is given by:
\bea A^{\EYM}_{n,2}(1,2,\ldots,n\,|\,p,q)&=&\sum_{\shuffle}(\eps_p \cdot Y_p)A^{\EYM}_{n+1,1}(1,\{2,\ldots,n-1\}\shuffle\{p\},n\,|\,q)\nonumber\\
&&+\sum_{\shuffle} (\eps_p\cdot F_q\cdot
Y_q)A_{n+2}^{\YM}(1,\{2,\ldots,n-1\}\shuffle\{q,p\},n)~.~~~\Label{G2-manifest-1}\eea
where we have used the field strength tensor:
\bea (F_q)^{\mu\nu}\equiv(k_q)^\mu(\epsilon_q)^{\nu}-(\epsilon_q)^\mu(k_q)^{\nu}\,.\Label{F-def}\eea

\item With three gravitons, the result is:
\bea
 A^{\EYM}_{n,3}(1,\ldots,n\,|\,p,q,r) &=&\sum_{\shuffle} (\eps_p \cdot Y_p)A^{\EYM}_{n+1,2}(1,\{2,\ldots,n-1\}\shuffle\{p\},n\,|\,q,r)\nonumber\\
&&+\sum_{\shuffle}(\eps_p\cdot F_q\cdot
Y_q) A^{\EYM}_{n+2,1}(1,\{2,\ldots,n-1\}\shuffle\{q,p\},n\,|\,r)\nonumber\\
&&+\sum_{\shuffle} (\eps_p\cdot F_r\cdot
Y_r)A^{\EYM}_{n+2,1}(1,\{2,\ldots,n-1\}\shuffle\{r,p\},n\,|\,q)\nonumber\\
&&+\sum_{\shuffle} (\eps_p\cdot F_q\cdot F_r\cdot
Y_r)A_{n+3}^{\YM}(1,\{2,\ldots,n-1\}\shuffle\{r,q,p\},n)\nonumber\\
&&+\sum_{\shuffle}(\eps_p\cdot F_r\cdot F_q\cdot
Y_q)A_{n+3}^{\YM}(1,\{2,\ldots,n-1\}\shuffle\{q,r,p\},n)\,.~~~\Label{threeGraRe}\eea

\end{itemize}
All the above relations are written in terms of amplitudes. However, using Eq.~\eref{eq:CHYintegration}, we can rewrite them in terms of CHY integrands, making the pattern even more transparent. With one graviton, we get:
\begin{equation}
\PT(12\ldots n)\,\Pf(\Psi_{p})=-\sum_{\shuffle} \left(\eps_p \cdot Y_p\right)\PT(1,\{2,\ldots,n-1\}\shuffle\{p\},n)\,,\Label{Tmu-0}
\end{equation}
which can be interpretted as turning the graviton into a gluon.
With two gravitons, we get:
\begin{align}
\PT(12\ldots n)\,\Pf(\Psi_{pq})&=\sum_{\shuffle}\left(\eps_p\cdot
Y_p\right)\PT(1,\{2,\ldots,n-1\}\shuffle\{p\},n)\,\Pf(\Psi_{q})\nonumber\\
&\quad-\sum_{\shuffle} \left(\eps_p\cdot F_q\cdot
Y_q\right)\PT(1,\{2,\ldots,n-1\}\shuffle\{q,p\},n)\,.\Label{G2-manifest-1-CHY}
\end{align}
With three gravitons, we get:
\begin{align}
\PT(12\ldots n)\,\Pf(\Psi_{pqr})&=-\sum_{\shuffle} \left(\eps_p \cdot Y_p\right)\PT(1,\{2,\ldots,n-1\}\shuffle\{p\},n)\,\Pf(\Psi_{qr})\nonumber\\
&\quad-\sum_{\shuffle}\left(\eps_p\cdot F_q\cdot
Y_q\right) \PT(1,\{2,\ldots,n-1\}\shuffle\{q,p\},n)\,\Pf(\Psi_{r})\nonumber\\
&\quad-\sum_{\shuffle} \left(\eps_p\cdot F_r\cdot
Y_r\right)\PT(1,\{2,\ldots,n-1\}\shuffle\{r,p\},n)\,\Pf(\Psi_{q})\nonumber\\
&\quad+\sum_{\shuffle} \left(\eps_p\cdot F_q\cdot F_r\cdot
Y_r\right)\PT(1,\{2,\ldots,n-1\}\shuffle\{r,q,p\},n)\nonumber\\
&\quad+\sum_{\shuffle}\left(\eps_p\cdot F_r\cdot F_q\cdot
Y_q\right)\PT(1,\{2,\ldots,n-1\}\shuffle\{q,r,p\},n)\,.\Label{threeGraRe-CHY}
\end{align}
Now it is clear that the general patterns observed in~\cite{Fu:2017uzt} indicates a pattern for CHY integrand expansion. In Section~\ref{sec:example} and~\ref{sec:recursive}, we will derive explicitly how this pattern appears. We note that the sign difference between the amplitude/integrand expansion originates from the phase factor in Eq.~\eqref{eq:CHYintegration}.

\section{Some examples}\label{sec:example}

In this section, we will use various on-shell and off-shell identities to show the pattern of CHY integrand expansion
with some examples, from which one can easily find the generalization. To make our discussion clear, we will carefully distinguish the identities we derive. The \emph{off-shell identities} are those that hold at the pure algebraic level. The \emph{on-shell identities} are true only when certain on-shell conditions
have been imposed. These on-shell conditions can be further divided into two types: the \emph{first type} is the momentum
conservation and the transverse condition of polarization vectors, and the \emph{second type} is that the $\sigma$'s in the integrands are solutions to the scattering equations. To manifest the distinction, we will use $=$ for off-shell identities, $\doteq$ for on-shell identities of the first type and $\cong$ for on-shell identities of the second type.


\subsection{One graviton}\label{sec:1h}

The relevant part of the EYM integrand $\mathcal{I}_{\text{EYM}}(12\ldots n\,|\,h_1)$ is
\begin{equation}
    \PT(12\ldots n)\,\Pf(\Psi_{h_1})\,,
\end{equation}
where $\Pf(\Psi_{h_1})$ can be easily calculated as:
\begin{equation}
    \Pf(\Psi_{h_1})=\Pf\left(\begin{array}{cc}
    0 & -C_{h_1h_1} \\
    C_{h_1h_1} & 0 \\
    \end{array}\right)=-C_{h_1h_1}=\sum_{i=1}^{n}\frac{\epsilon_{h_1}\cdot k_i}{\sigma_{h_1i}}\,.
\end{equation}
To move on, we need the following very useful identity:
\begin{equation}
    \frac{\PT(12\ldots n)}{\sigma_{ih_1}}=\PT(12\ldots i,h_1,i+1\ldots n)
    +\frac{\PT(12\ldots n)}{\sigma_{i+1,h_1}}\,.\Label{eq:insertsigma}
\end{equation}
The proof is very simple:
\bea \frac{\PT(12\ldots n)}{\sigma_{ih_1}}
    & = & \PT(12\ldots i,h_1,i+1\ldots
    n)\left(\frac{\sigma_{h_1,i+1}}{\sigma_{i,i+1}}\right)\,.\nn
&= & \PT(12\ldots i,h_1,i+1\ldots n)+\frac{\sigma_{h_1i}}{\sigma_{i,i+1}}
\PT(12\ldots i,h_1,i+1\ldots n)\nonumber\\
    &=& \PT(12\ldots i,h_1,i+1\ldots n)+\frac{\PT(12\ldots n)}
    {\sigma_{i+1,h_1}}\,.\Label{Cii-recursive}
\eea
where we have used $\sigma_{h_1,i+1}=\sigma_{h_1i}+\sigma_{i,i+1}$
to go from the first line to second. We want to remark that many papers use instead
$C_{ii}=\sum_{j\neq t}\epsilon_{i}\cdot k_j\left(\frac{\sigma_{jt}}{\sigma_{ji}\sigma_{it}}\right)$. Although this equivalent form has the right $SL(2,\mathbb{C})$ weight
for the $i$-th node, we find that our form of $C_{ii}$ is more convenient for our purpose, for example, the recursive manipulation in Eq.~\eqref{Cii-recursive}. Furthermore, we
emphasize that Eq.~\eqref{eq:insertsigma} holds at the algebraic level
for arbitrary set of $\sigma$'s, not necessarily those of the
solutions to the scattering equations. In this sense, it is an
off-shell identity.

Now with the help of Eq.~\eqref{eq:insertsigma}, we find that
\bea  & & C_{h_1h_1}\PT(12\ldots n)=\PT(12\ldots
n)\left(\frac{\epsilon_{h_1}\cdot k_1}
{\sigma_{1h_1}}+\ldots+\frac{\epsilon_{h_1}\cdot
k_n}{\sigma_{nh_1}}\right)\nn
&=& \sum_{\shuffle}\left(\epsilon_{h_1}\cdot Y_{h_1}\right)
    \PT(1,\{2\ldots n-1\}\shuffle\{h_1\},n)+ \eps_{h_1}\cdot \left(\sum_{i=1}^n k_i\right)
   \frac{\PT(12\ldots n)}{\sigma_{n,h_1}}\,\nn
&\doteq& \sum_{\shuffle}\left(\epsilon_{h_1}\cdot Y_{h_1}\right)
    \PT(1,\{2\ldots n-1\}\shuffle\{h_1\},n)\,,\Label{eq:C1PT}
\eea
where we have used the on-shell condition $\eps_{h_1}\cdot
(\sum_{i=1}^n k_i)=-\eps_{h_1}\cdot k_{h_1}=0$ at the second line. Therefore, Eq.~\eqref{eq:C1PT} holds under the on-shell condition of the first type, as indicated by the $\doteq$ sign in the last line. With the help of Eq.~\eqref{eq:C1PT}, we arrive at
\begin{align}
\Pf(\Psi_{h_1})\,\PT(12\ldots n)\doteq -\sum_{\shuffle}\left(\epsilon_{h_1}\cdot Y_{h_1}\right)\,\PT(1,\{2\ldots n-1\}\shuffle\{h_1\},n)\,,
\Label{eq:h1result}\end{align}
or, in terms of the amplitudes,
\begin{equation}
    A_{n,1}^{\rm EYM}(12\ldots n\,|\,h_1)=\sum_{\shuffle}\left(\epsilon_{h_1}\cdot Y_{h_1}\right)A_{n+1}^{\rm YM}(1,\{2\ldots n-1\}\shuffle\{h_1\},n)\,.
\Label{eq:1hamp}
\end{equation}
 This is nothing but~\eref{Tmu-0}. The sign change is accounted by the phase factor of Eq.~\eqref{eq:CHYintegration}.

For future use, we define the following level-zero-current CHY integrand $T^\mu$
\be T^{\mu}\left[1,\{2\ldots n-1\}\shuffle\{h_1\ldots
h_i\},n\,|\varnothing\right]\equiv\sum_{\shuffle}
Y^{\mu}_{h_1}\,\PT(1,\{2\ldots n-1\}\shuffle\{h_1\ldots h_i\},n)\,,\Label{eq:T1} \ee
%
Like the usual CHY integrands, $T^\mu$ has weight two for all legs, so it is well defined even without imposing the scattering equations. However, $T^\mu$ has some special properties. First, its parameters are divided to two parts. The first part  contains two ordered sets $\{2\ldots n-1\}$ and $\{h_1\ldots h_i\}$. The second part contains an unordered set. For Eq.~\eqref{eq:T1} it is the empty set $\varnothing$, and this is the reason why Eq.~\eqref{eq:T1} is called the level-zero-current. Secondly, Eq.~\eqref{eq:T1} contains the Lorentz indices through $Y_{h_1}^\mu$. It is worth to notice that $h_1$ is the first element of the ordered list $\{h_1,\ldots, h_i\}$.  With this new notation, Eq.~\eqref{eq:h1result} can be written as:
\begin{equation}
    \Pf(\Psi_{h_1})\,\PT(12\ldots n)\doteq
    -\epsilon_{h_1}\cdot T\left[1,\{2\ldots n-1\}\shuffle\{h_1\},n|\,\varnothing\right]\,.
\end{equation}
%

\subsection{Two gravitons}\label{sec:2h}

The relevant part of the EYM integrand $\mathcal{I}_{\text{EYM}}(12\ldots n\,|\,h_2h_1)$ is given by
\begin{equation}
\PT(12\ldots n)\,\Pf(\Psi_{h_2h_1})\,.\Label{eq:h2integrand}
\end{equation}
To continue, we expand $\Pf(\Psi_{h_2h_1})$ as
\bea
    \Pf(\Psi_{h_2h_1})&= & \Pf\left(\begin{array}{cccc}
    0 & A_{h_2h_1} & -C_{h_2h_2} & -C_{h_1h_2} \\
    A_{h_1h_2} & 0 & -C_{h_2h_1} & -C_{h_1h_1} \\
    C_{h_2h_2} & C_{h_2h_1} & 0 & B_{h_2h_1} \\
    C_{h_1h_2} & C_{h_1h_1} & B_{h_1h_2} & 0 \\
    \end{array}\right)\nonumber\\
    &= & C_{h_2h_2}\Pf(\Psi^{n+2}_{h_1})-C_{h_2h_1}\Pf\left(\begin{array}{cc}
    0 & -C_{h_1h_2} \\
    C_{h_1h_2} & 0 \\
    \end{array}\right)+B_{h_2h_1}\Pf\left(\begin{array}{cc}
    0 & A_{h_2h_1} \\
    A_{h_1h_2} & 0 \\
    \end{array}\right)\,,\Label{Phi-2g-exp}
\eea
where the superscript $n+2$ in $\Psi_{h_1}^{n+2}$ is to remind us that we are now working with the $(n+2)$-particle kinematics, although the form of $\Psi_{h_1}^{n+2}$ resembles the $\Psi_{h_1}$ in Section~\ref{sec:1h}. Similar to Eq.~\eqref{eq:C1PT}, when $C_{h_2h_2}$ hits $\PT(12\ldots n)$, we get:
\begin{align}
 C_{h_2h_2}\PT(12\ldots n)&=\PT(12\ldots n)\left(\frac{\epsilon_{h_2}\cdot k_1}
 {\sigma_{1h_2}}+\ldots+\frac{\epsilon_{h_2}\cdot k_n}{\sigma_{nh_2}}
 +\frac{\epsilon_{h_2}\cdot k_{h_1}}{\sigma_{h_1h_2}}\right)\nn
&=\sum_{\shuffle}\left(\epsilon_{h_2}\cdot
Y_{h_2}\right)\PT(1,\{2\ldots n-1\} \shuffle\{h_2\},n)\nn
&\quad +
\frac{\eps_{h_2}\cdot \sum_{i=1}^n k_i}{\sigma_{n h_2}}\PT(12\ldots
n)+\PT(12\ldots n)\frac{\epsilon_{h_2}\cdot
k_{h_1}}{\sigma_{h_1h_2}}\nn
& \doteq \sum_{\shuffle}\left(\epsilon_{h_2}\cdot
Y_{h_2}\right)\PT(1,\{2\ldots n-1\} \shuffle\{h_2\},n)
-\frac{\sigma_{nh_1}}{\sigma_{nh_2}}\,C_{h_2h_1}\PT(12\ldots n)\,.
\Label{eq:C2PT}\end{align}
Now we give some remarks to this equation. First, when using Eq.~\eqref{eq:insertsigma} to insert $h_2$ into gluons, only the gluon momenta contribute, so we get $\eps_{h_2}\cdot Y_{h_2}$. Secondly, going from the second equality to the third, we have used the momentum conservation and $\eps_{h_2}\cdot k_{h_2}=0$, as indicated by the $\doteq$ symbol. Thirdly, the above manipulation is, in fact, the action of \emph{changing one graviton to gluon}, which results in the first term of Eq.~\eqref{eq:C2PT}. However, in the presence of other gravitons, we need some corrections [such as the second term in Eq.~\eqref{eq:C2PT}]. As we will see shortly, the recursive structure is hidden in these corrections.

The relation \eref{eq:C2PT} can be easily generalized to the cases with more gravitons as
\begin{align}
    C_{h_ih_i}\PT(12\ldots n)&\doteq\sum_{\shuffle}\left(\epsilon_{h_i}\cdot Y_{h_i}\right)
    \PT(1,\{2\ldots n-1\}\shuffle\{h_i\},n)-\sum_{\substack{j=1 \\ j\neq i}}^{m}\frac{\sigma_{nh_j}C_{h_ih_j}}{\sigma_{nh_i}}\PT(12\ldots n)\,,
\Label{eq:CmPT}\end{align}
for any $h_i\in\mathsf{H}=\{h_mh_{m-1}\ldots h_1\}$. We will use \eref{eq:CmPT} in later computations again and again. A further observation is that, when we make the replacement $\epsilon_{h_i}\rightarrow k_{h_i}$ and impose the scattering equations, we will make $C_{h_i h_i}\rightarrow 0$ at the left hand side, such that Eq.~\eqref{eq:CmPT} leads to:
\begin{equation}
    \sum_{\shuffle}\left(k_{h_i}\cdot Y_{h_i}\right)\PT(1,\{2\ldots n-1\}\shuffle\{h_i\},n) \cong\sum_{j=1}^{m}\frac{\sigma_{nh_j}A_{h_ih_j}}{\sigma_{nh_i}}\PT(12\ldots n)\,.
\Label{eq:k1result}\end{equation}
We have used the $\cong$ symbol to emphasize that this equation holds only under the on-shell condition of the second type, namely, localizing $\sigma$'s to be the solutions to the scattering equations.

Now plugging Eq.~\eqref{eq:C2PT} back into Eq.~\eqref{eq:h2integrand}, we can rewrite Eq.~\eqref{eq:h2integrand} into:
\begin{align}
    \PT(12\ldots n)\,\Pf(\Psi_{h_2h_1})&\doteq \sum_{\shuffle}\left(\epsilon_{h_2}\cdot
    Y_{h_2}\right)\Pf(\Psi^{n+2}_{h_1})\,\PT(1,\{2\ldots n-1\}\shuffle\{h_2\},n)\nonumber\\
    &\quad -C_{h_2h_1}\left[\frac{\sigma_{nh_1}}{\sigma_{nh_2}}\Pf(\Psi^{n+2}_{h_1})+\Pf\left(\begin{array}{cc}
        0 & -C_{h_1h_2} \\
        C_{h_1h_2} & 0 \\
    \end{array}\right)\right]\PT(12\ldots n)\nonumber\\
    &\quad+B_{h_2h_1}\Pf\left(\begin{array}{cc}
    0 & A_{h_2h_1} \\
    A_{h_1h_2} & 0 \\
    \end{array}\right)\PT(12\ldots n)\,.
\Label{eq:h2step1}\end{align}
The first line of Eq.~\eqref{eq:h2step1} is now in its final form, which gives:
\begin{equation*}
    -\sum_{\shuffle}\left(\epsilon_{h_2}\cdot Y_{h_2}\right)A^{\rm EYM}_{n+1,1}(1,\{2\ldots n-1\}\shuffle\{h_2\},n\,|\,h_1)
\end{equation*}
after the CHY integration. Now we consider the second line of
\eref{eq:h2step1}, which contains two terms.  The first term contains the factor
\begin{equation*}
    \Pf(\Psi^{n+2}_{h_1})\,\PT(12\ldots n)\,.
\end{equation*}
It would just be the one-graviton case solved in Section~\ref{sec:1h}, had $h_2$ joined in the Parke-Taylor factor. This observation guides us to turn $\PT(12\ldots n)$ into $\PT(12\ldots nh_2)$ by inserting the factor $\sigma_{n1}/(\sigma_{nh_2}\sigma_{h_21})$, such that the previous result can be used directly: 
\bea
    && \frac{\sigma_{nh_1}}{\sigma_{nh_2}}\,\Pf(\Psi^{n+2}_{h_1})\,\PT(12\ldots n)=\frac{\sigma_{nh_1}\sigma_{h_21}}{\sigma_{n1}}\,\Pf(\Psi^{n+2}_{h_1})\,\PT(12\ldots nh_2)\nonumber\\
    &\doteq & -\frac{\sigma_{nh_1}\sigma_{h_21}}{\sigma_{n1}}\sum_{\shuffle}
    \left(\epsilon_{h_1}\cdot Y_{h_1}\right)\PT(1,\{2\ldots n\}\shuffle\{h_1\},h_2)\nonumber\\
    &\doteq & -\frac{\sigma_{nh_1}}{\sigma_{nh_2}}\sum_{\shuffle}\left(\epsilon_{h_1}\cdot
    Y_{h_1}\right)\PT(1,\{2\ldots n-1\}\shuffle\{h_1\},n)+C_{h_1h_2}\PT(12\ldots n)\,.
\Label{eq:Pf1PT}\eea
To go from the second line to third line, we have separated the orderings in the shuffle $\{2\ldots n\}\shuffle\{h_1\}$ into
two groups, in which: (1) $n$ being the last element;
(2) $h_1$ being the last element. The first group gives the first term in the third line of \eref{eq:Pf1PT}.
For the second group, we get
\bean -\frac{\sigma_{nh_1}\sigma_{h_21}}{\sigma_{n1}}
   \left( \epsilon_{h_1}\cdot \sum_{i=1}^n k_i\right)\PT(12\ldots n h_1 h_2)= \frac{\eps_{h_1}\cdot k_{h_2}}{\sigma_{h_1 h_2}}\,\PT(12\ldots n )=C_{h_1h_2}\,\PT(12\ldots n)\,.\eean
Putting \eref{eq:Pf1PT} back into the second row of Eq.~\eqref{eq:h2step1}, we get:
\bea
    & & -C_{h_2h_1}\left[\frac{\sigma_{nh_1}}{\sigma_{nh_2}}\Pf(\Psi^{n+2}_{h_1})-C_{h_1h_2}\right]\PT(12\ldots n)\nonumber\\
    &\doteq & \frac{\sigma_{nh_1}(\epsilon_{h_2}\cdot k_{h_1})}{\sigma_{nh_2}\sigma_{h_2h_1}}
    \sum_{\shuffle}\left(\epsilon_{h_1}\cdot Y_{h_1}\right)\PT(1,\{2\ldots n-1\}\shuffle\{h_1\},n)\nonumber\\
    &\doteq & -(\epsilon_{h_2}\cdot k_{h_1})\sum_{\shuffle}\left(\epsilon_{h_1}\cdot
    Y_{h_1}\right)\PT(1,\{2\ldots n-1\}\shuffle\{h_1h_2\},n)\,,\Label{eq:2shuffle}\eea
which is its final form. To derive the last line of \eref{eq:2shuffle}, we have to use an insertion relation, which can be derived as follows:
\begin{align}
    \frac{\sigma_{h_1n}}{\sigma_{h_1h_2}\sigma_{h_2n}}&=\left(\int_{\sigma_{h_1}}^{\sigma_{i+1}}+\int_{\sigma_{i+1}}^{\sigma_{i+2}}+\ldots+\int_{\sigma_{n-1}}^{\sigma_n}\right)\frac{d\sigma}{(\sigma-\sigma_{h_2})^2}\nonumber\\
    &=\frac{\sigma_{h_1,i+1}}{\sigma_{h_1h_2}\sigma_{h_2,i+1}}+\frac{\sigma_{i+1,i+2}}{\sigma_{i+1,h_2}\sigma_{h_2,i+2}}+\ldots+\frac{\sigma_{n-1,n}}{\sigma_{n-1,h_2}\sigma_{h_2n}}\,,
\Label{eq:separate}\end{align}
We want to emphasize that Eq.~\eqref{eq:separate}  is an off-shell identity. This manipulation can be generalized to
\begin{align}
\label{eq:attach}
    &\quad\frac{\sigma_{h_in}}{\sigma_{h_ih_{j}}\sigma_{h_{j}n}}\sum_{\shuffle}Y_{h_1}^{\mu}\,\PT(1,\{2\ldots n-1\}\shuffle\{h_1\ldots h_i\},n)\nonumber\\
    &=\sum_{\shuffle}Y^{\mu}_{h_1}\,\PT(1,\{2\ldots n-1\}\shuffle\{h_1\ldots h_ih_j\},n)\,,
\end{align}
where the set $\{h_1\ldots h_i\}$ does not contain $h_j$. This identity will be very useful later.

Now we come to calculate the last row of Eq.~\eqref{eq:h2step1}.
Actually, the result can be immediately read out from our previous
calculations. After we replace $\epsilon_{h_1}$ by $ k_{h_1}$ into
Eq.~\eqref{eq:Pf1PT},  we get zero in the left hand side, since $\Pf(\Psi^{n+2}_{h_1})\to 0$ under this replacement, if $\sigma$'s are the solutions to the scattering equations. Thus we obtain:
\begin{equation}
    A_{h_1h_2}\,\PT(12\ldots n)\cong\frac{\sigma_{nh_1}}{\sigma_{nh_2}}\sum_{\shuffle}
    \left(k_{h_1}\cdot Y_{h_1}\right)\PT(1,\{2\ldots n-1\}\shuffle\{h_1\},n)\,,
\end{equation}
 With this equation, the last row of Eq.~\eqref{eq:h2step1} becomes
\begin{align}
    -B_{h_2h_1}A_{h_1h_2}\,\PT(12\ldots n)
    &\cong(\epsilon_{h_2}\cdot\epsilon_{h_1})\,\frac{\sigma_{h_1n}}
    {\sigma_{h_1h_2}\sigma_{h_2n}}\sum_{\shuffle}\left(k_{h_1}\cdot Y_{h_1}\right)\PT(1,\{2\ldots n-1\}\shuffle\{h_1\},n)\nonumber\\
    &\cong(\epsilon_{h_2}\cdot\epsilon_{h_1})\sum_{\shuffle}\left(k_{h_1}\cdot Y_{h_1}\right)\PT(1,\{2\ldots n-1\}\shuffle\{h_1h_2\},n)\,.
\Label{eq:Bh2}\end{align}
In fact, the relation between the second line and the third line of Eq.~\eqref{eq:h2step1} can be more easily seen
by noticing that $C_{h_2h_1}|_{k_{h_1}\to \eps_{h_1}}  =   B_{h_2h_1}$, and consequently,
\begin{equation}
  \left.\left[\frac{\sigma_{nh_1}}{\sigma_{nh_2}}\Pf(\Psi^{n+2}_{h_1})+\Pf\left(\begin{array}{cc}
        0 & -C_{h_1h_2} \\
        C_{h_1h_2} & 0 \\
    \end{array}\right)\right]\right|_{\epsilon_{h_1}\rightarrow k_{h_1}}\cong\Pf\left(\begin{array}{cc}
    0 & A_{h_2h_1} \\
    A_{h_1h_2} & 0 \\
    \end{array}\right)\,.\Label{F-appearing}
\end{equation}
Putting together Eq.~\eqref{eq:2shuffle} and \eqref{eq:Bh2}, we can write down the final result for Eq.~\eqref{eq:h2step1}:
\bea
    \PT(12\ldots n)\,\Pf(\Psi_{h_2h_1})&\cong & \sum_{\shuffle}\left(\epsilon_{h_2}\cdot Y_{h_2}\right)\Pf(\Psi^{n+2}_{h_1})\,\PT(1,\{2\ldots n-1\}\shuffle\{h_2\},n)\nonumber\\
    & & -\sum_{\shuffle}\left(\epsilon_{h_2}\cdot F_{h_1}\cdot Y_{h_1}\right)\PT(1,\{2\ldots n-1\}\shuffle\{h_1h_2\},n)\,,
\Label{eq:h2result}\eea
where we have used the definition \eref{F-def} to combine the second and third line of Eq.~\eqref{eq:h2step1}.
Namely, we can expand the two-graviton EYM integrand in terms of one-graviton integrands and pure gluon ones.
Eq.~\eqref{eq:h2result} leads to the following amplitude relation:
\bea
    A_{n,2}^{\rm EYM}(12\ldots n\,|\,h_2h_1)&= & \sum_{\shuffle}\left(\epsilon_{h_2}\cdot Y_{h_2}\right)A_{n+1,1}^{\rm EYM}(1,\{2\ldots n-1\}\shuffle\{h_2\},n\,|\,h_1)\nonumber\\
    & & +\sum_{\shuffle}\left(\epsilon_{h_2}\cdot F_{h_1}\cdot Y_{h_1}\right)A_{n+2}^{\rm YM}(1,\{2\ldots n-1\}\shuffle\{h_1h_2\},n)~~~
\Label{eq:2hamp}\eea
The sign difference between Eq.~\eqref{eq:2hamp} and Eq.~\eqref{eq:h2result} is correctly adjusted according to the phase factor in the CHY integration~\eqref{eq:CHYintegration} for EYM amplitudes.


Similar to what we have done in Eq.~\eqref{eq:T1}, we define the following level-one-current CHY integrand
\begin{align}
    T^{\mu}\left[1,\{2\ldots n-1\}\shuffle\{h_2\ldots h_i\},n |\{h_1\}\right]&=\sum_{\shuffle} Y^{\mu}_{h_2}\,\Pf(\Psi^{n+i}_{h_1})\,\PT(1,\{2\ldots n-1\}\shuffle\{h_2\ldots h_i\},n)\nonumber\\
    &\quad -(F_{h_1})^{\mu\nu}T_{\nu}\left[1,\{2\ldots n-1\}\shuffle\{h_1h_2\ldots h_i\},n|\,\varnothing\right]\,.
  \Label{eq:T2}
\end{align}
It is easy to see that after using Eq.~\eref{eq:T1} at the second line, we can re-express Eq.~\eqref{eq:h2result} as:
\begin{equation}
    \PT(12\ldots n)\,\Pf(\Psi_{h_2h_1}) \cong
    \epsilon_{h_2}\cdot T\left[1,\{2\ldots n-1\}\shuffle\{h_2\},n |\{h_1\}\right]\,,
\end{equation}
which holds for an arbitrary number of gluons.

\subsection{Three gravitons}\label{sec:3h}
As the first nontrivial example, we come to expand the three-graviton EYM integrand:
\begin{equation*}
    \PT(12\ldots n)\,\Pf(\Psi_{h_3h_2h_1})\,.
\end{equation*}
We first expand $\Pf(\Psi_{h_3h_2h_1})$ into the following form:\footnote{This expansion comes from the standard Laplace formula Eq.~\eqref{eq:laplace}, with some additional re-arrangements in the algebraic minors so that
	those associated with $C$ and $B$ elements share the same sign.}
\begin{align}
    \Pf(\Psi_{h_3h_2h_1})&=\Pf\left(\begin{array}{cccccc}
    0 & A_{h_3h_2} & A_{h_3h_1} & -C_{h_3h_3} & -C_{h_2h_3} & -C_{h_1h_3} \\
    A_{h_2h_3} & 0 & A_{h_2h_1} & -C_{h_3h_2} & -C_{h_2h_2} & -C_{h_1h_2} \\
    A_{h_1h_3} & A_{h_1h_2} & 0 & -C_{h_3h_1} & -C_{h_2h_1} & -C_{h_1h_1} \\
    C_{h_3h_3} & C_{h_3h_2} & C_{h_3h_1} & 0 & B_{h_3h_2} & B_{h_3h_1} \\
    C_{h_2h_3} & C_{h_2h_2} & C_{h_2h_1} & B_{h_2h_3} & 0 & B_{h_2h_1} \\
    C_{h_1h_3} & C_{h_1h_2} & C_{h_1h_1} & B_{h_1h_3} & B_{h_1h_2} & 0 \\
    \end{array}\right)\nonumber\\
    &=-C_{h_3h_3}\Pf(\Psi^{n+3}_{h_2h_1})+C_{h_3h_2}\Pf\left[\psi(32|1)\right]+C_{h_3h_1}\Pf\left[\psi(31|2)\right]\nonumber\\
    &\quad-B_{h_3h_2}\Pf\left[\O{\psi}(32|1)\right]-B_{h_3h_1}\Pf\left[\O{\psi}(31|2)\right]\,.
\Label{eq:3hPsi}\end{align}
In this equation, the matrix $\psi(32|1)$ is given by:
\begin{equation}
    \psi(32|1)\equiv\left(\begin{array}{cccc}
    0 & A_{h_3h_1} & -C_{h_2h_3} & -C_{h_1h_3} \\
    A_{h_1h_3} & 0 & -C_{h_2h_1} & -C_{h_1h_1} \\
    C_{h_2h_3} & C_{h_2h_1} & 0 & B_{h_2h_1} \\
    C_{h_1h_3} & C_{h_1h_1} & B_{h_1h_2} & 0 \\
    \end{array}\right)\,.
\end{equation}
Namely, $\psi(32|1)$ is obtained from $\Psi_{h_3h_2h_1}$ by deleting the row and column intersected at $\pm C_{h_3h_2}$. Then $\psi(31|2)$ is obtained from $\psi(32|1)$ by the exchange $h_2\leftrightarrow h_1$. On the other hand, the matrix
$\O\psi(32|1)$ is obtained from $\psi(32|1)$ by the replacement $\epsilon_{h_2}\rightarrow k_{h_2}$, and similarly for ${\O\psi}(31|2)$. Now using Eq.~\eqref{eq:CmPT} for the first term of \eqref{eq:3hPsi} to turn the graviton $h_3$ into a gluon, we arrive at:
\begin{align}
    \PT(12\ldots n)\,\Pf(\Psi_{h_3h_2h_1})&\doteq -\sum_{\shuffle}\left(\epsilon_{h_3}\cdot Y_{h_3}\right)\Pf(\Psi^{n+3}_{h_2h_1})\,\PT(1,\{2\ldots n-1\}\shuffle\{h_3\},n)\nonumber\\
    &\quad+C_{h_3h_2}\left[\frac{\sigma_{nh_2}}{\sigma_{nh_3}}\,\Pf(\Psi^{n+3}_{h_2h_1})+\Pf\left[\psi(32|1)\right]\right]\PT(12\ldots n)\nonumber\\
    &\quad+C_{h_3h_1}\left[\frac{\sigma_{nh_1}}{\sigma_{nh_3}}\,\Pf(\Psi^{n+3}_{h_2h_1})+\Pf\left[\psi(31|2)\right]\right]\PT(12\ldots n)\nonumber\\
    &\quad-B_{h_3h_2}\Pf\left[\O{\psi}(32|1)\right]\PT(12\ldots n)\nonumber\\
    &\quad-B_{h_3h_1}\Pf\left[\O{\psi}(31|2)\right]\PT(12\ldots n)\,.
\Label{eq:h3step1}\end{align}
At this moment, after comparing Eq.~\eqref{eq:h3step1} with
\eqref{eq:h2step1}, one can see a pattern starting to emerge. Another thing we want to point out
is the relative sign difference between the first line of
\eref{eq:h3step1} and \eref{eq:h2step1}.

In Eq.~\eqref{eq:h3step1}, the first row is already in our desired form. We then focus on the second row, while the third row can be obtained by simply exchanging $h_1$ and $h_2$. We would
like to reuse Eq.~\eqref{eq:h2result} to expand
$\Pf(\Psi^{n+3}_{h_2h_1})\,\PT(12\ldots n)$. To do so, we again use
the trick of inserting $h_3$ into the Parke-Taylor factor. Then using Eq.~\eqref{eq:h2result}, we obtain
\bea
    \Pf(\Psi^{n+3}_{h_2h_1})\,\PT(12\ldots nh_3)&\cong &
    \sum_{\shuffle}\left(\epsilon_{h_2}\cdot Y_{h_2}\right)\Pf(\Psi^{n+3}_{h_1})\,
    \PT(1,\{2\ldots n\}\shuffle\{h_2\},h_3)\nonumber\\
    && -\sum_{\shuffle}\left(\epsilon_{h_2}\cdot F_{h_1}\cdot Y_{h_1}\right)\PT(1,\{2\ldots n\}
    \shuffle\{h_1h_2\},h_3)\,.\Label{eq:h3step2}\eea
{In this equation, we treat $h_3$ as a gluon, playing the same role as $n$ in Eq.~\eqref{eq:h2result}. In the first line of Eq.~\eqref{eq:h3step2}, the shuffle product can be divided into two groups: one with $n$ near $h_3$ and the other with $h_2$ near $h_3$: 
\begin{equation*}
	\{2\ldots n\}\shuffle\{h_2\}=\{2\ldots n-1\}\shuffle\{h_2\}n+\{2\ldots nh_2\}\,.
\end{equation*}}
After this separation, the first line of Eq.~\eqref{eq:h3step2} becomes:
\begin{align}
  &\quad\sum_{\shuffle}\left(\epsilon_{h_2}\cdot Y_{h_2}\right)\Pf(\Psi^{n+3}_{h_1})\,\PT(1,\{2\ldots n\}\shuffle\{h_2\},h_3)\nonumber\\
  &\doteq\sum_{\shuffle}\left(\epsilon_{h_2}\cdot Y_{h_2}\right)\Pf(\Psi^{n+3}_{h_1})\,\PT(1,\{2\ldots n-1\}\shuffle\{h_2\},n,h_3)\nonumber\\
  &\quad-\left[\epsilon_{h_2}\cdot (k_{h_1}+k_{h_3})\right]\Pf(\Psi^{n+3}_{h_1})\,\PT(12\ldots nh_2h_3)\nn
  &\doteq \sum_{\shuffle}\left(\epsilon_{h_2}\cdot Y_{h_2}\right)\Pf(\Psi^{n+3}_{h_1})\,
\PT(1,\{2\ldots n-1\}\shuffle\{h_2\},n,h_3)\nonumber\\
  &\quad+\left[\epsilon_{h_2}\cdot(k_{h_1}+k_{h_3})\right]\sum_{\shuffle}\left(\epsilon_{h_1}\cdot
X_{h_1}\right)\PT(1,\{2\ldots nh_2\}\shuffle\{h_1\},h_3)\,,
\Label{eq:firstrow}
\end{align}
where the momentum conservation  $\sum_{i=1}^nk_i=-k_{h_1}-k_{h_2}-k_{h_3}$ has been used to derive the expression after the first equal sign. {The final expression is then obtained by acting $\Pf(\Psi^{n+3}_{h_1})$ onto $\PT(12\ldots nh_2h_3)$ using Eq.~\eqref{eq:h1result}.} In the last line of Eq.~\eqref{eq:firstrow}, we have introduced a new symbol $X_{h_1}$. This is because when we apply \eqref{eq:h1result} here, $h_1$ will view $h_2$ as a gluon, so we use the $X_{h_1}$ instead of $Y_{h_1}$ to emphasize this difference. From now on, we will use $X_{i}$ to stand for the sum of momenta of all the legs at the left hand side of leg $i$ in the color ordering. With this explanation, we can expand the last line of \eref{eq:firstrow} as:
{\begin{align}
	\sum_{\shuffle}\left(\epsilon_{h_1}\cdot X_{h_1}\right)\PT(1,\{2\ldots nh_2\}\shuffle\{h_1\},h_3)&\doteq\sum_{\shuffle}\left(\epsilon_{h_1}\cdot Y_{h_1}\right)\PT(1,\{2\ldots n-1\}\shuffle\{h_1\},n,h_2,h_3)\nonumber\\
	&\quad-\epsilon_{h_1}\cdot\left(k_{h_2}+k_{h_3}\right)\PT(12\ldots nh_1h_2h_3)\nonumber\\
	&\quad-\left(\epsilon_{h_1}\cdot k_{h_3}\right)\PT(12\ldots nh_2h_1h_3)\,,
\end{align}}
so that the final result of Eq.~\eqref{eq:firstrow} is:
\begin{align}
  &\quad\sum_{\shuffle}\left(\epsilon_{h_2}\cdot Y_{h_2}\right)\Pf(\Psi^{n+3}_{h_1})\,\PT(1,\{2\ldots n\}\shuffle\{h_2\},h_3)\nonumber\\
  &\doteq\sum_{\shuffle}\left(\epsilon_{h_2}\cdot
Y_{h_2}\right)\Pf(\Psi^{n+3}_{h_1})\,
\PT(1,\{2\ldots n-1\}\shuffle\{h_2\},n,h_3)\nonumber\\
  &\quad+\left[\epsilon_{h_2}\cdot (k_{h_1}+k_{h_3})\right]
\sum_{\shuffle}\left(\epsilon_{h_1}\cdot
Y_{h_1}\right)\PT(1,\{2\ldots n-1\}\shuffle\{h_1\},n,h_2,h_3)\nn
  &\quad-\left[\epsilon_{h_2}\cdot (k_{h_1}+k_{h_3})\right]\left[\eps_{h_1}\cdot (k_{h_2}+k_{h_3})\right]\PT(12...n h_1 h_2
h_3)\nonumber\\
  &\quad-\left[\epsilon_{h_2}\cdot (k_{h_1}+k_{h_3})\right](\eps_{h_1}\cdot k_{h_3})\,\PT(12...n  h_2 h_1
h_3)\,.\Label{eq:firstrow-2}
\end{align}
Now we move to the second row of Eq.~\eqref{eq:h3step2}. Again we separate the shuffle product into three groups
\begin{equation*}
  \{2\ldots n\}\shuffle\{h_1h_2\}=\{2\ldots n-1\}\shuffle\{h_1h_2\}n+\{2\ldots n-1\}\shuffle\{h_1\}nh_2+\{2\ldots nh_1h_2\}\,.
\end{equation*}
This separation leads to:
\begin{align}
  &\quad-\sum_{\shuffle}\left(\epsilon_{h_2}\cdot F_{h_1}\cdot Y_{h_1}\right)\PT(1,\{2\ldots n\}\shuffle\{h_1h_2\},h_3)\nn
  &=-\sum_{\shuffle}\left(\epsilon_{h_2}\cdot F_{h_1}\cdot Y_{h_1}\right)\PT(1,\{2\ldots n-1\}\shuffle\{h_1h_2\},n,h_3)\nn
  &\quad-\sum_{\shuffle}\left(\epsilon_{h_2}\cdot F_{h_1}\cdot Y_{h_1}\right)\PT(1,\{2\ldots n-1\}\shuffle\{h_1\},n, h_2,h_3)\nn
  &\quad+\left[\epsilon_{h_2}\cdot F_{h_1}\cdot (k_{h_2}+k_{h_3})\right]\PT(12\ldots
    nh_1h_2h_3) \Label{eq:secondrow1}
\end{align}
Summing up Eq.~\eqref{eq:firstrow-2} and \eqref{eq:secondrow1}, we find that after a few algebras, Eq.~\eqref{eq:h3step2} becomes:
\begin{subequations}
\label{eq:secondrow}
\begin{align} 
\Pf(\Psi^{n+3}_{h_2h_1})\,\PT(12\ldots nh_3)&\cong\sum_{\shuffle}\left(\epsilon_{h_2}\cdot
Y_{h_2}\right)\Pf(\Psi^{n+3}_{h_1})\, \PT(1,\{2\ldots
n-1\}\shuffle\{h_2\},n,h_3) \nn
&\quad  -\sum_{\shuffle}\left(\epsilon_{h_2}\cdot F_{h_1}\cdot
Y_{h_1}\right)\PT(1,\{2\ldots n-1\}
    \shuffle\{h_1h_2\},n,h_3)\nn
\label{eq:a}
&\quad +(\epsilon_{h_2}\cdot k_{h_3})
\sum_{\shuffle}\left(\epsilon_{h_1}\cdot
Y_{h_1}\right)\PT(1,\{2\ldots n-1\}\shuffle\{h_1\},n,h_2,h_3)\\
\label{eq:b}
&\quad -\left(\epsilon_{h_2}\cdot k_{h_3}\right) \eps_{h_1}\cdot
\left(k_{h_2}+k_{h_3}\right)\PT(12\ldots
nh_1h_2h_3)\\
\label{eq:c}
&\quad-\left(\epsilon_{h_2}\cdot k_{h_3}\right)\left(\epsilon_{h_1}\cdot k_{h_3}\right)\PT(12\ldots nh_2h_1h_3)\\
\label{eq:d}
&\quad -\left(\epsilon_{h_2}\cdot k_{h_1}\right) \left(\eps_{h_1}\cdot
k_{h_3}\right) \PT(12...n  h_2 h_1 h_3)\\
\label{eq:e}
&\quad + (\epsilon_{h_2}\cdot \eps_{h_1})
\sum_{\shuffle}\left(k_{h_1}\cdot Y_{h_1}\right)\PT(1,\{2\ldots
n-1\}\shuffle\{h_1\},n,h_2,h_3)\\
\label{eq:f}
&\quad - \left(\epsilon_{h_2}\cdot \eps_{h_1}\right) k_{h_1}\cdot
\left(k_{h_2}+k_{h_3}\right)\PT(12\ldots
 nh_1h_2h_3)\,.
\end{align}
\end{subequations}
{The first two lines of this equation are in our desired form, while the rest of them look very messy at the first glance. Very remarkably, they can actually be organized into $\Pf[\psi(32|1)]$. To show this, we first use the identity:
\begin{align}
\label{eq:h1gn+3}
\Pf(\Psi^{n+3}_{h_1})\,\PT(12\ldots nh_2h_3)&\doteq-\sum_{\shuffle}\left(\epsilon_{h_1}\cdot Y_{h_1}\right)\PT(1,\{2\ldots n-1\}\shuffle\{h_1\},n,h_2,h_3)\nonumber\\
&\quad+\epsilon_{h_1}\cdot\left(k_{h_2}+k_{h_3}\right)\PT(12\ldots nh_1h_2h_3)\nonumber\\
&\quad+\left(\epsilon_{h_1}\cdot k_{h_3}\right)\PT(12\ldots nh_2h_1h_3)
\end{align}
to group Eq.~\eqref{eq:a}, \eqref{eq:b} and \eqref{eq:c} into:
\begin{align}
\label{eq:abc}
\eqref{eq:a}+\eqref{eq:b}+\eqref{eq:c}&\doteq-\left(\epsilon_{h_2}\cdot k_{h_3}\right)\Pf(\Psi^{n+3}_{h_1})\,\PT(12\ldots nh_2h_3)\nonumber\\
&=-\left(\frac{\sigma_{n1}}{\sigma_{nh_2}\sigma_{h_31}}\right)C_{h_2h_3}\Pf(\Psi_{h_1}^{n+3})\,\PT(12\ldots n)\,.
\end{align}
Next, by extracting $h_2$ and $h_3$ out of the Parke-Taylor factor, we can write Eq.~\eqref{eq:d} as:
\begin{align}
\label{eq:d2}
\eqref{eq:d}&=-\left(\frac{\sigma_{n1}}{\sigma_{nh_2}\sigma_{h_31}}\right)C_{h_2h_1}C_{h_1h_3}\PT(12\ldots n)\nonumber\\
&=\left(\frac{\sigma_{n1}}{\sigma_{nh_2}\sigma_{h_31}}\right)C_{h_2h_1}\,\Pf\left(\begin{array}{cc} 0 & -C_{h_1h_3} \\ C_{h_1h_3} & 0 \\ \end{array}\right)\PT(12\ldots n)\,.
\end{align}
As the last step, we replace the $\epsilon_{h_1}$ by $k_{h_1}$ in Eq.~\eqref{eq:h1gn+3}, which leads to the on-shell identity:
\begin{align}
\sum_{\shuffle}\left(k_{h_1}\cdot Y_{h_1}\right)\PT(1,\{2\ldots n-1\}\shuffle\{h_1\},n,h_2,h_3)&\cong k_{h_1}\cdot\left(k_{h_2}+k_{h_3}\right)\PT(12\ldots nh_1h_2h_3)\nonumber\\
&\quad +(k_{h_1}\cdot k_{h_3})\,\PT(12\ldots nh_2h_1h_3)\,.
\end{align}
With the help of this result, Eq.~\eqref{eq:e} and \eqref{eq:f} combine into:
\begin{align}
\label{eq:ef}
\eqref{eq:e}+\eqref{eq:f}&\cong-\left(\frac{\sigma_{n1}}{\sigma_{nh_2}\sigma_{h_31}}\right)B_{h_2h_1}A_{h_3h_1}\,\PT(12\ldots n)\nonumber\\
&=-\left(\frac{\sigma_{n1}}{\sigma_{nh_2}\sigma_{h_31}}\right)B_{h_2h_1}\,\Pf\left(\begin{array}{cc} 0 & A_{h_3h_1}
\\ A_{h_1h_3} & 0 \end{array}\right)\PT(12\ldots n)\,.
\end{align}
Now collecting Eq.~\eqref{eq:abc}, \eqref{eq:d2} and \eqref{eq:ef}, one can show with some algebras that the last six lines of Eq.~\eqref{eq:secondrow} can be combined to:}
\begin{align}
&\quad\eqref{eq:a}+\eqref{eq:b}+\eqref{eq:c}+\eqref{eq:d}+\eqref{eq:e}+\eqref{eq:f}\nonumber\\
&\cong-\frac{\sigma_{n1}}{\sigma_{nh_2}\sigma_{h_31}}\left[C_{h_2h_3}\,\PT(\Psi_{h_1}^{n+3})-C_{h_2h_1}\,\Pf\left(\begin{array}{cc} 0 & -C_{h_1h_3} \\ C_{h_1h_3} & 0 \\ \end{array}\right)+B_{h_2h_1}\,\Pf\left(\begin{array}{cc} 0 & A_{h_3h_1}
      \\ A_{h_1h_3} & 0 \end{array}\right)\right]\PT(12\ldots n)\nn
  &=-\left(\frac{\sigma_{n1}}{\sigma_{nh_2}\sigma_{h_31}}\right)\Pf[\psi(32|1)]~\PT(12\ldots n)\,.
\end{align}
Namely, we have proved the following identity:
\bea
    & &\Pf(\Psi_{h_2h_1}^{n+3})\,\PT(12\ldots nh_3)+\frac{\sigma_{n1}}
    {\sigma_{nh_2}\sigma_{h_31}}\,\Pf[\psi(32|1)]~\PT(12\ldots n)\nonumber\\
   &\cong &
\sum_{\shuffle}\left(\epsilon_{h_2}\cdot
Y_{h_2}\right)\Pf(\Psi^{n+3}_{h_1})\, \PT(1,\{2\ldots
n-1\}\shuffle\{h_2\},n,h_3) \nn
& &  -\sum_{\shuffle}\left(\epsilon_{h_2}\cdot F_{h_1}\cdot
Y_{h_1}\right)\PT(1,\{2\ldots n-1\}
\shuffle\{h_1h_2\},n,h_3)\,.\Label{secondrow-simplifed}\eea
With all these preparations, we can now return to tackle the second row of Eq.~\eqref{eq:h3step1}. Using Eq.~\eqref{secondrow-simplifed}, we find that:
\bea  & & C_{h_3
h_2}\left[\frac{\sigma_{nh_2}}{\sigma_{nh_3}}\,\Pf(\Psi^{n+3}_{h_2h_1})\,
+\Pf[\psi(32|1)]\right] ~\PT(12\ldots n)\nn
& = & C_{h_3 h_2}\left(\frac{\sigma_{nh_2}\sigma_{h_31}}{\sigma_{n1}}\right)\left[\Pf(\Psi_{h_2h_1}^{n+3})\,\PT(12\ldots nh_3)
+\frac{\sigma_{n1}}{\sigma_{nh_2}\sigma_{h_31}}\,\Pf[\psi(32|1)]~\PT(12\ldots n)\right]\nonumber\\
&\cong &-(\eps_{h_3}\cdot k_{h_2})\left(\frac{\sigma_{h_2 n}}{\sigma_{h_2h_3}\sigma_{ h_3 n }}\right)\left[\sum_{\shuffle}\left(\epsilon_{h_2}\cdot
Y_{h_2}\right)\Pf(\Psi^{n+3}_{h_1})\, \PT(1,\{2\ldots
n-1\}\shuffle\{h_2\},n) \right.\nn
& & \left. -\sum_{\shuffle}\left(\epsilon_{h_2}\cdot F_{h_1}\cdot
Y_{h_1}\right)\PT(1,\{2\ldots n-1\}
    \shuffle\{h_1h_2\},n)\right]\,.
\label{eq:Pf2PTn}\eea
Finally, using the insertion identity \eref{eq:separate}, we get:
\begin{align}
  &\quad C_{h_3h_2}\left[\frac{\sigma_{nh_2}}{\sigma_{nh_3}}\,\Pf(\Psi^{n+3}_{h_2h_1})\,
+\Pf[\psi(32|1)]\right]\PT(12\ldots n) \nn
  &\cong -(\eps_{h_3}\cdot k_{h_2})\left[\sum_{\shuffle}\left(\epsilon_{h_2}\cdot Y_{h_2}\right)\Pf(\Psi^{n+3}_{h_1})\, \PT(1,\{2\ldots n-1\}\shuffle\{h_2 h_3\},n)\right.\nn
  &\quad\qquad\qquad\qquad\left. -\sum_{\shuffle}\left(\epsilon_{h_2}\cdot F_{h_1}\cdot
Y_{h_1}\right)\PT(1,\{2\ldots n-1\}
    \shuffle\{h_1h_2 h_3\},n)\right]\,.
  \Label{eq:Pf2PTn-1}
\end{align}
Next, noting that the fourth line of \eqref{eq:h3step1} can be obtained from the
second line by the exchange $\eps_{h_2}\leftrightarrow k_{h_2}$
[just like Eq.~\eqref{F-appearing}], we can write down the result directly from Eq.~\eqref{eq:Pf2PTn-1} as:
\begin{align}
  &\quad B_{h_3h_1}\Pf\left[\overline{\psi}(32|1)\right]\PT(12\ldots n)\nn
  &\cong -(\eps_{h_3}\cdot \epsilon_{h_2})\left[\sum_{\shuffle}\left(k_{h_2}\cdot Y_{h_2}\right)\Pf(\Psi^{n+3}_{h_1})\,\PT(1,\{2\ldots n-1\}\shuffle\{h_2 h_3\},n)\right.\nn
  &\quad\qquad\qquad\qquad\left.-\sum_{\shuffle}\left(k_{h_2}\cdot F_{h_1}\cdot
Y_{h_1}\right)\PT(1,\{2\ldots n-1\}
    \shuffle\{h_1h_2 h_3\},n)\right]\,.\Label{eq:BPf2PTn-1}
\end{align}
Thus combining \eqref{eq:Pf2PTn-1} and \eqref{eq:BPf2PTn-1} together, we arrive at:
\begin{align}
  &\quad\;C_{h_3h_2}\left[\frac{\sigma_{nh_2}}{\sigma_{nh_3}}\,\Pf(\Psi^{n+3}_{h_2h_1})+\Pf[\psi(32|1)]\right]\PT(12\ldots n)-B_{h_3h_1}\Pf\left[\overline{\psi}(32|1)\right]\PT(12\ldots n)\nn
  &\cong-\sum_{\shuffle}\left(\epsilon_{h_3}\cdot F_{h_2}\cdot
Y_{h_2}\right)\Pf(\Psi^{n+3}_{h_1})\, \PT(1,\{2\ldots
n-1\}\shuffle\{h_2 h_3\},n) \nn
&\quad +\sum_{\shuffle}\left(\epsilon_{h_3}\cdot F_{h_2}\cdot
F_{h_1}\cdot Y_{h_1}\right)\PT(1,\{2\ldots n-1\}
\shuffle\{h_1h_2 h_3\},n)\,.
\end{align}
Then the third and fifth line of Eq.~\eqref{eq:h3step1} can be obtained by replacing $h_2\leftrightarrow h_1$ in the above equation.
Finally, putting everything altogether, we reach the following result:
\bea \PT(12\ldots n)\,\Pf(\Psi_{h_3h_2h_1})&\doteq &
-\sum_{\shuffle}\left(\epsilon_{h_3}\cdot
Y_{h_3}\right)\Pf(\Psi^{n+3}_{h_2h_1})\, \PT(1,\{2\ldots
n-1\}\shuffle\{h_3\},n)\nn
& & -\sum_{\shuffle}\left(\epsilon_{h_3}\cdot F_{h_2}\cdot
Y_{h_2}\right)\Pf(\Psi^{n+3}_{h_1})\, \PT(1,\{2\ldots
n-1\}\shuffle\{h_2 h_3\},n) \nn
& & -\sum_{\shuffle}\left(\epsilon_{h_3}\cdot F_{h_1}\cdot
Y_{h_2}\right)\Pf(\Psi^{n+3}_{h_2})\, \PT(1,\{2\ldots
n-1\}\shuffle\{h_1 h_3\},n) \nn
& &  +\sum_{\shuffle}\left(\epsilon_{h_3}\cdot F_{h_2}\cdot
F_{h_1}\cdot Y_{h_1}\right)\PT(1,\{2\ldots n-1\}
    \shuffle\{h_1h_2 h_3\},n)\nn
& &  +\sum_{\shuffle}\left(\epsilon_{h_3}\cdot F_{h_1}\cdot
F_{h_2}\cdot Y_{h_2}\right)\PT(1,\{2\ldots n-1\}
    \shuffle\{h_2h_1 h_3\},n)\Label{eq:h3result} \eea
It is wroth noting that after considering the relative signs encoded in Eq.~\eqref{eq:CHYintegration}, this result indeed gives Eq.~\eqref{threeGraRe} after the CHY integration.

Eq.~\eqref{eq:h3result} demonstrates a very clear recursive structure, while we can make it more compact by defining the level-$(m-1)$-current CHY integrand [like the one given in Eq.~\eqref{eq:T2}]:
\bea
    & &  T^{\mu}\left[1,\{2\ldots n-1\}\shuffle\{h_m\ldots h_i\},n\,|\{ h_1,...,h_{m-1}\}
    \right]\nn
    &\equiv & \sum_{\shuffle} Y_{h_m}^{\mu}\Pf(\Psi^{n+i}_{h_{1}\ldots h_{m-1}})
    \,\PT(1,\{2\ldots n-1\}\shuffle\{h_m\ldots h_i\},n)\nn
    & &+(-1)^{m-1}\sum_{s=1}^{m-1}(F_{h_s})^{\mu\nu}
    T_{\nu}\left[1,\{2\ldots n-1\}\shuffle\{h_sh_m\ldots h_i\},n\,|\{h_{m-1}
    \ldots\slashed{h}_s\ldots h_{m-1}\}\right]\,.
\Label{eq:recursion-1}\eea
This is a recursive definition, and one can move down by reducing
the number of gravitons in the right list until it becomes the empty set. Then Eq.~\eqref{eq:T1} terminates the whole recursive construction. We can use $T^{\mu}$ to rewrite Eq.~\eqref{eq:h3result} as:
\bea
    \Pf(\Psi_{h_3h_2h_1})\,\PT(12\ldots n)\cong
    (-1)^{3}\epsilon_{h_3}\cdot T\left[1,\{2\ldots n-1\}\shuffle\{h_3\},n\,|\,\{h_2h_1\}\right]\,.\Label{3g-T-exp}
\eea
This three-graviton example makes the recursive pattern very clear. Our next job is of course to prove that it works for arbitrary number of gravitons. To get more confidence and insight, we do one more example with four gravitons. The techniques used in this calculation can be very easily generalized to prove the generic case, presented in Section~\ref{sec:recursive}.

\subsection{Four gravitons}\label{sec:4h}

As the last example, we expand the Pfaffian part of the four-graviton
integrand
 $   \Pf(\Psi_{h_4h_3h_2h_1})\,\PT(12\ldots n) $
as follows:
\bea
     \Pf(\Psi_{h_4h_3h_2h_1})&=&C_{h_4h_4}\Pf(\Psi^{n+4}_{h_3h_2h_1})\nonumber\\
    && -C_{h_4h_3}\Pf\left[\psi(43|21)\right]-C_{h_4h_2}\Pf\left[\psi(42|31)\right]-C_{h_4h_1}\Pf\left[\psi(41|32)\right]\nonumber\\
    & & +B_{h_4h_3}\Pf\left[\bar\psi(43|21)\right]+B_{h_4h_2}\Pf\left[\bar\psi(42|31)\right]+B_{h_4h_1}\Pf\left[\bar\psi(41|32)\right]\,.
\Label{eq:4hPsi}\eea
where  $\Psi_{h_3h_2h_1}^{n+4}$ has a very similar form to that of
Eq.~\eqref{eq:3hPsi}. However, now the $\sigma$'s satisfy instead the $n+4$ point
scattering equations, while in the diagonal $C$ elements, the summation is over all the $n+4$ particles. The matrix $\psi(43|21)$ is obtained from $\Psi_{h_4h_3h_2h_1}$ by deleting the column and row intersected at $\pm C_{h_4h_3}$:
\begin{equation}
    \psi(43|21)=\left(\begin{array}{cccccc}
    0 & A_{h_4h_2} & A_{h_4h_1} & -C_{h_3h_4} & -C_{h_2h_4} & -C_{h_1h_4} \\
    A_{h_2h_4} & 0 & A_{h_2h_1} & -C_{h_3h_2} & -C_{h_2h_2} & -C_{h_1h_2} \\
    A_{h_1h_4} & A_{h_1h_2} & 0 & -C_{h_3h_1} & -C_{h_2h_1} & -C_{h_1h_1} \\
    C_{h_3h_4} & C_{h_3h_2} & C_{h_3h_1} & 0 & B_{h_3h_2} & B_{h_3h_1} \\
    C_{h_2h_4} & C_{h_2h_2} & C_{h_2h_1} & B_{h_2h_3} & 0 & B_{h_2h_1} \\
    C_{h_1h_4} & C_{h_1h_2} & C_{h_1h_1} & B_{h_1h_3} & B_{h_1h_2} & 0 \\
    \end{array}\right)\,,
\end{equation}
while $\psi(42|31)$ and $\psi(41|32)$ are obtained from this matrix by exchanging the index $h_3$ with $h_2$ and $h_1$ respectively. The matrix $\O\psi(\ldots s|\ldots)$ is obtained from the corresponding $\psi(\ldots s|\ldots)$ by the replacement $\epsilon_{h_s}\rightarrow k_{h_s}$.

We start from the first line of Eq.~\eqref{eq:4hPsi}. When combining $C_{h_4h_4}$ with the Parke-Taylor factor, we can turn the graviton $h_4$ into a gluon with some extra pieces:
\begin{align}
    C_{h_4h_4}\,\PT(12\ldots n)&\doteq\sum_{\shuffle}\left(\epsilon_{h_4}\cdot Y_{h_4}\right)\PT(1,\{2\ldots n-1\}\shuffle\{h_4\},n)-\sum_{s=1}^{3}\frac{\sigma_{nh_s}C_{h_4h_s}}{\sigma_{nh_4}}\,\PT(12\ldots n)\,.
\end{align}
Putting it back to \eqref{eq:4hPsi} and reorganizing, we get:
\begin{align}
&\quad\,\Pf(\Psi_{h_4h_3h_2h_1})\,\PT(12\ldots n)\nn
&\doteq\sum_{\shuffle}\left(\epsilon_{h_4}\cdot Y_{h_4}\right)\PT(1,\{2\ldots n-1\}\shuffle\{h_4\},n)\Pf(\Psi^{n+4}_{h_3h_2h_1})\nn
    &\quad+\left\{  -C_{h_4h_3}\left[\frac{\sigma_{nh_3}}{\sigma_{nh_4}}\,\Pf(\Psi_{h_3h_2h_1}^{n+4})+\Pf\left[\psi(43|21)\right]\right]+B_{h_4h_3}\,\Pf\left[\bar\psi(43|21)\right]\right\}\PT(12\ldots n)\nn
    &\quad+(h_3\leftrightarrow h_2)+(h_3\leftrightarrow h_1)\,.\Label{eq:h4step1}\end{align}
We only focus on the case $s=3$, as explicitly shown in the second line from the bottom in this equation. The other two terms in the last line can be obtained from the trivial replacements $h_3\leftrightarrow h_2$ and $h_3\leftrightarrow h_1$. First, according to Eq.~\eqref{3g-T-exp}, we have:
\begin{align}
\label{eq:Psi3PTn+1}
    \frac{\sigma_{nh_3}}{\sigma_{nh_4}}\Pf(\Psi_{h_3h_2h_1}^{n+4})\,\PT(12\ldots n)&=\left(\frac{\sigma_{nh_3}\sigma_{h_41}}{\sigma_{n1}}\right)\Pf(\Psi_{h_3h_2h_1}^{n+4})\,\PT(12\ldots nh_4)\nonumber\\
    &\cong-\left(\frac{\sigma_{nh_3}\sigma_{h_41}}{\sigma_{n1}}\right)\epsilon_{h_3}\cdot T\left[1,\{2\ldots n\}\shuffle\{h_3\},h_4\,|\,\{h_2h_1\}\right]\,.
\end{align}
The expansion of $T^{\mu}\left[1,\{2\ldots n\}\shuffle\{h_3\},h_4\,|\,\{h_2h_1\}\right]$ involves the shuffle products of the form:
\begin{equation*}
    \{\ldots n\}\shuffle\{\star\star\star\,h_3\}=\{\ldots n-1\}\shuffle\{\star\star\star\, h_3\}\,n+\{\ldots n\}\shuffle\{\star\star\star\}\,h_3\,,
\end{equation*}
such that it can be separated in the following way, according to the definition \eqref{eq:recursion-1}:
\begin{align}
\label{eq:h4step2}
    T^{\mu}\left[1,\{2\ldots n\}\shuffle\{h_3\},h_4\,|\,\{h_2h_1\}\right]&\doteq T^{\mu}\left[1,\{2\ldots n-1\}\shuffle\{h_3\},n,h_4\,|\,\{h_2h_1\}\right]\nonumber\\
    &\quad-\left[\sum_{s=1}^{4}(k_{h_s})^{\mu}\right]\Pf(\Psi_{h_2h_1}^{n+4})\,\PT(12\ldots nh_3h_4)\nonumber\\
    &\quad+(F_{h_2})^{\mu\nu}T_{\nu}\left[1,\{2\ldots n\}\shuffle\{h_2\},h_3,h_4\,|\,\{h_1\}\right]\nonumber\\
    &\quad+(F_{h_1})^{\mu\nu}T_{\nu}\left[1,\{2\ldots n\}\shuffle\{h_1\},h_3,h_4\,|\,\{h_2\}\right]
\end{align}
Now using the following identities proved in the three-graviton example:\footnote{The matrix $\psi(432|1)$ $[\psi(431|2)]$ is obtained from $\psi(43|21)$ by deleting the row and column intersected at $\pm C_{h_3h_2}$ $(\pm C_{h_3h_1})$.}
\bea
    -\epsilon_{h_2}\cdot T\left[1,\{2\ldots
    n\}\shuffle\{h_2h_4\},h_3\,|\,\{h_1\}\right]
    & \cong & \frac{1}{\sigma_{h_4h_2}}\left[\frac{\sigma_{h_3h_2}}{\sigma_{h_3h_4}}\,
    \Pf(\Psi_{h_2h_1}^{n+4})+\Pf\left[\psi(432|1)\right]\right]\PT(12\ldots nh_3)\nn
    -k_{h_2}\cdot T\left[1,\{2\ldots
    n\}\shuffle\{h_2h_4\},h_3\,|\,\{h_1\}\right]
    & \cong & \frac{1}{\sigma_{h_4h_2}}\,\Pf\left[\bar\psi(432|1)\right]\PT(12\ldots nh_3)\nn
    -\epsilon_{h_1}\cdot T\left[1,\{2\ldots n\}\shuffle\{h_1h_4\},h_3\,|\,\{h_2\}\right]&
    \cong & \frac{1}{\sigma_{h_4h_1}}\left[\frac{\sigma_{h_3h_1}}{\sigma_{h_3h_4}}\,
    \Pf(\Psi_{h_2h_1}^{n+4})+\Pf\left[\psi(431|2)\right]\right]\PT(12\ldots nh_3)\nn
      -k_{h_1}\cdot T\left[1,\{2\ldots n\}\shuffle\{h_1h_4\},h_3\,|\,\{h_2\}\right]& \cong &\frac{1}{\sigma_{h_4h_1}}\,\Pf\left[\bar\psi(431|2)\right]\PT(12\ldots nh_3)
      \,,
\eea
 and Eq.~\eqref{eq:attach}, we can derive that:
\begin{align}
    \epsilon_{h_2}\cdot T
    \left[1,\{2\ldots n\}\shuffle\{h_2\},h_3,h_4\,|\,\{h_1\}\right]&\cong \Pf(\Psi_{h_2h_1}^{n+4})\,\PT(12\ldots nh_3h_4)\nn
    &\quad+\frac{\sigma_{h_31}}{\sigma_{h_3h_2}
    \sigma_{h_41}}\,\Pf\left[\psi(432|1)\right]\PT(12\ldots nh_3) \nn
    k_{h_2}\cdot T\left[1,\{2\ldots n\}\shuffle\{h_2\},h_3,h_4\,|\,\{h_1\}\right]&\cong\frac{\sigma_{h_31}}{\sigma_{h_3h_2}\sigma_{h_41}}\Pf\left[\bar\psi(432|1)\right]
    \PT(12\ldots nh_3)\,,.
\end{align}
%
After we replace the $T$'s in the last two lines of Eq.~\eqref{eq:h4step2} by the above eqaution, and dot the entire Eq.~\eqref{eq:h4step2} with $\epsilon_{h_3}$, it becomes:
\begin{align}
    &\quad\;\epsilon_{h_3}\cdot T\left[1,\{2\ldots n\}\shuffle\{h_3\},h_4\,|\,\{h_2h_1\}\right]\nonumber\\
    &\cong\epsilon_{h_3}\cdot T\left[1,\{2\ldots n-1\}\shuffle\{h_3\},n,h_4\,|\,\{h_2h_1\}\right]\nonumber\\
    &\quad+\frac{\sigma_{h_31}}{\sigma_{h_41}}\Big[-C_{h_3h_4}\,\Pf(\Psi_{h_2h_1}^{n+4})+C_{h_3h_2}\,\Pf\left[\psi(432|1)\right]+C_{h_3h_1}\,\Pf\left[\psi(431|2)\right]\nonumber\\
    &\qquad\qquad\quad-B_{h_3h_2}\,\Pf\left[\bar\psi(432|1)\right]-B_{h_3h_1}\,\Pf\left[\bar\psi(431|2)\right]\Big]\PT(12\ldots nh_3)\nonumber\\
    &\cong\epsilon_{h_3}\cdot T\left[1,\{2\ldots n-1\}\shuffle\{h_3\},n,h_4\,|\,\{h_2h_1\}\right]+\frac{\sigma_{h_31}}{\sigma_{h_41}}\,\Pf\left[\psi(43|21)\right]\PT(12\ldots nh_3)\,.
\end{align}
If we put them back into Eq.~\eqref{eq:Psi3PTn+1} and then Eq.~\eqref{eq:h4step1}, we get:
\begin{align}
\label{eq:psiT4}
&\quad C_{h_4h_3}\left[\frac{\sigma_{nh_3}}{\sigma_{nh_4}}\,\Pf(\Psi_{h_3h_2h_1}^{n+4})+\Pf\left[\psi(43|21)\right]\right]\PT(12\ldots n)\nonumber\\
&\cong\left(\epsilon_{h_4}\cdot k_{h_3}\right)\epsilon_{h_3}\cdot T\left[1,\{2\ldots n-1\}\shuffle\{h_3h_4\},n\,|\,\{h_2h_1\}\right]\,.
\end{align}
If we make the replacement $\epsilon_{h_3}\rightarrow k_{h_3}$ above, we then get:
\begin{equation}
\label{eq:psiU4}
    \frac{1}{\sigma_{h_4h_3}}\,\Pf\left[\bar\psi(43|21)\right]\cong k_{h_3}\cdot T\left[1,\{2\ldots n-1\}\shuffle\{h_3h_4\},n\,|\,\{h_2h_1\}\right]\,.
\end{equation}
{The $(h_3\leftrightarrow h_2)$ and $(h_3\leftrightarrow h_1)$ term of Eq.~\eqref{eq:h4step1} can then be obtained by making the corresponding exchange in Eq.~\eqref{eq:psiT4} and \eqref{eq:psiU4}.} The final result of Eq.~\eqref{eq:h4step1} is thus:
\begin{align}
\Pf(\Psi_{h_4h_3h_2h_1})\,\PT(12\ldots n)&\cong\sum_{\shuffle}\left(\epsilon_{h_4}\cdot Y_{h_4}\right)\Pf(\Psi_{h_3h_2h_1}^{n+4})\,\PT(1,\{2\ldots n-1\}\shuffle\{h_4\},n)\nonumber\\
&\quad-\epsilon_{h_4}\cdot F_{h_3}\cdot T\left[1,\{2\ldots n-1\}\shuffle\{h_3h_4\},n|\{h_2h_1\}\right]\nonumber\\
&\quad-\epsilon_{h_4}\cdot F_{h_2}\cdot T\left[1,\{2\ldots n-1\}\shuffle\{h_2h_4\},n|\{h_3h_1\}\right]\nonumber\\
&\quad-\epsilon_{h_4}\cdot F_{h_1}\cdot T\left[1,\{2\ldots n-1\}\shuffle\{h_1h_4\},n|\{h_3h_2\}\right]\,.
\end{align}
Finally, using the definition \eref{eq:recursion-1}, one can express the four-graviton EYM integrand as:
\begin{equation}
    \Pf(\Psi_{h_4h_3h_2h_1})\,\PT(12\ldots n)\cong(-1)^{4}\,\epsilon_{h_4}\cdot T\left[1,\{2\ldots n-1\}\shuffle\{h_4\},n\,|\,\{h_3h_2h_1\}\right]\,.
\end{equation}
The recursive nature of this construction is now clear,  and we are going to prove the generic recursive relation for expanding EYM integrands with an arbitrary number of gravitons.

\section{General recursive relation}\label{sec:recursive}

The calculations in Section~\ref{sec:example}, especially the cases of three and four gravitons, suggest us to construct the recursive definition \eref{eq:recursion-1}, which we recall here as:
\begin{align}
\label{eq:recursion}
    &\quad T^{\mu}\left[1,\{2\ldots n-1\}\shuffle\{h_m\ldots h_i\},n\,|\,\{h_{m-1}\ldots h_1\}\right]\nonumber\\
    &=\sum_{\shuffle} Y_{h_m}^{\mu}\Pf(\Psi^{n+i}_{h_{m-1}\ldots h_1})\,\PT(1,\{2\ldots n-1\}\shuffle\{h_m\ldots h_i\},n)\nonumber\\
    &\quad+(-1)^{m-1}\sum_{s=1}^{m-1}(F_{h_s})^{\mu\nu}
    T_{\nu}\left[1,\{2\ldots n-1\}\shuffle\{h_sh_m\ldots h_i\},n\,|\,\{h_{m-1}\ldots\slashed{h}_s\ldots h_1\}\right]\,.
\end{align}
This recursion finally lands on $T\left[\ldots|\varnothing\right]$ defined in
Eq.~\eqref{eq:T1},  for arbitrary $i\geqslant m$. This definition
gives a well defined weight two integrand for $n+i$ particles. One
crucial point we want to mention is that $$T^{\mu}\left[1,\{2\ldots
n-1\}\shuffle\{h_m\ldots h_i\},n\,|\,\{h_{m-1}\ldots h_1\}\right]$$ has the
manifest gauge invariance for the gravitons $\{h_1h_2...h_{m-1}\}$. In this section, we are going to prove by induction a very
important result:
\bea
    \Pf(\Psi_{\mathsf{H}})\,\PT(12\ldots n)\cong (-1)^{m}\epsilon_{h_m}\cdot T
    \left[1,\{2\ldots n-1\}\shuffle\{h_m\},n\,|\,\{h_{m-1}\ldots h_1\}\right]
\Label{eq:hmresult}\eea
for $\mathsf{H}=\{h_mh_{m-1}\ldots h_1\}$. In this representation,
$h_m$ is singled out as the row along which we choose to expand
the Pfaffian. Because of this choice, the gauge invariance for $\{h_1\ldots h_{m-1}\}$ is manifest, but not for $h_m$. Only after we fully expand it to pure YM amplitudes, can we check the gauge invariance of $h_m$ with various KK and generalized BCJ relations~\cite{Bern:2008qj}.

The $m=1$, $2$, $3$ and $4$ cases have been explicitly checked  in
the previous section. Now suppose that Eq.~\eqref{eq:hmresult}
holds for $p$ gravitons with $p\leqslant m-1$, as our induction
assumption, then we move to the case of $m$ gravitons. The $2m\times
2m$ matrix $\Psi_{\mathsf{H}}$ has the following form:
\begin{equation}
    \Psi_{\mathsf{H}}=\left(\begin{array}{cccccc}
        0 & A_{h_mh_{m-1}} & \cdots & -C_{h_mh_m} & -C_{h_{m-1}h_m} & \cdots \\
        A_{h_{m-1}h_m} & 0 & \cdots & -C_{h_mh_{m-1}} & -C_{h_{m-1}h_{m-1}} & \cdots \\
        \vdots & \vdots & & \vdots & \vdots & \\
        C_{h_mh_m} & C_{h_mh_{m-1}} & \cdots & 0 & B_{h_mh_{m-1}} & \cdots \\
        C_{h_{m-1}h_m} & C_{h_{m-1}h_{m-1}} & \cdots & B_{h_{m-1}h_m} & 0 & \cdots \\
        \vdots & \vdots & & \vdots & \vdots & \\
    \end{array}\right)\,.
\end{equation}
To calculate the Pfaffian, we expand along the row starting with
$C_{h_mh_m}$ [which is the $(m+1)$-th row]:
\bea
    (-1)^{m}\Pf(\Psi_{\mathsf{H}})=C_{h_mh_m}\Pf(\Psi_{h_{m-1}\ldots h_1}^{n+m})&-&\sum_{s=1}^{m-1}C_{h_mh_s}\Pf[\psi(m,s|m-1\ldots\slashed{s}\ldots 1)]\nonumber\\
    &+&\sum_{s=1}^{m-1}B_{h_mh_s}\Pf[\bar{\psi}(m,s|m-1\ldots\slashed{s}\ldots 1)]\,.
~~~~\Label{eq:PsiH}\eea
In this expansion, $\Psi_{h_{m-1}\ldots h_1}^{n+m}$ is obtained from
$\Psi_{\mathsf{H}}$ by deleting the rows and columns intersected at
the $\pm C_{h_mh_m}$. We note that $\Psi_{h_{m-1}\ldots h_1}^{n+m}$
is part of the EYM integrand with $m-1$ gravitons, while $h_m$ has
been turned into a gluon. The matrix $\psi(m,{m-1}\,|\,{m-2}\ldots 1)$
is obtained from $\Psi_{\mathsf{H}}$ by deleting the rows and
columns intersected at the $\pm C_{h_mh_{m-1}}$, while the other
$\psi(m,s\,|\,{m-1}\ldots\slashed{s}\ldots 1)$ can be generated by
exchanging $h_{m-1}$ and $h_s$. In the last row of
Eq.~\eqref{eq:PsiH}, we have:
\begin{equation*}
    \O{\psi}(m,s\,|\,{m-1}\ldots\slashed{s}\ldots 1)=\left.\psi(m,s\,|\,{m-1}\ldots\slashed{s}\ldots 1)\right|_{\epsilon_{h_s}\rightarrow k_{h_s}}\,.
\end{equation*}
After multiplying Eq.~\eqref{eq:PsiH} with the Parke-Taylor factor $\PT(12\ldots n)$, the first term yields:
\begin{align}
     C_{h_mh_m}\Pf(\Psi_{h_{m-1}\ldots h_1}^{n+m})\PT(12\ldots n)&\doteq\sum_{\shuffle}\left(\epsilon_{h_m}\cdot Y_{h_m}\right)\Pf(\Psi_{h_{m-1}\ldots h_1}^{n+m})\,\PT(1,\{2\ldots n-1\}\shuffle\{h_m\},n)\nonumber\\
    &\quad -\sum_{s=1}^{m-1}\frac{\sigma_{nh_s}C_{h_mh_s}}{\sigma_{nh_m}}\,\Pf(\Psi_{h_{m-1}\ldots h_1}^{n+m})\,\PT(12\ldots n)\,,
\end{align}
 with the help of Eq.~\eqref{eq:CmPT}. Then from Eq.~\eqref{eq:PsiH}, we get:
\begin{align}
\label{eq:hmstep1}
    &\quad\left(-1\right)^{m}\Pf(\Psi_{\mathsf{H}})\,\PT(12\ldots n)\nonumber\\
    &\doteq\sum_{\shuffle}\left(\epsilon_{h_m}\cdot Y_{h_m}\right)\Pf(\Psi_{h_{m-1}\ldots h_1}^{n+m})\,\PT(1,\{2\ldots n-1\}\shuffle\{h_m\},n)\nonumber\\
    &\quad -\sum_{s=1}^{m-1}\frac{\epsilon_{h_m}\cdot k_{h_s}}{\sigma_{h_mh_s}}\left[\frac{\sigma_{nh_s}}{\sigma_{nh_m}}\,\Pf(\Psi_{h_{m-1}\ldots h_1}^{n+m})+\Pf[\psi(m,s\,|\,{m-1}\ldots\slashed{s}\ldots 1)]\right]\PT(12\ldots n)\nonumber\\
    &\quad +\sum_{s=1}^{m-1}\frac{\epsilon_{h_m}\cdot\epsilon_{h_s}}{\sigma_{h_mh_s}}\,\Pf[\bar{\psi}(m,s\,|\,{m-1}\ldots\slashed{s}\ldots 1)]\,\PT(12\ldots n)\,.
\end{align}
Our proof completes if we can prove the following two identities:
\begin{align}
\label{eq:Tm-1}
    &\quad\;\frac{1}{\sigma_{h_mh_s}}\left[\frac{\sigma_{nh_s}}{\sigma_{nh_m}}\,\Pf(\Psi_{h_{m-1}\ldots h_1}^{n+m})+\Pf[\psi(m,s\,|\,{m-1}\ldots\slashed{s}\ldots 1)]\right]\PT(12\ldots n)\nonumber\\
    &\cong(-1)^{m}\epsilon_{h_s}\cdot T\left[1,\{2\ldots n-1\}\shuffle\{h_sh_m\},n\,|\,\{h_{m-1}\ldots\slashed{h}_s\ldots h_1\}\right]\,,\\
\label{eq:Um-1}
    &\quad\;\frac{1}{\sigma_{h_mh_s}}\,\Pf[\bar{\psi}(m,s\,|\,{m-1}\ldots\slashed{s}\ldots 1)]\,\PT(12\ldots n)\nonumber\\
    &\cong(-1)^{m}k_{h_s}\cdot T\left[1,\{2\ldots n-1\}\shuffle\{h_sh_m\},n\,|\,\{h_{m-1}\ldots\slashed{h}_s\ldots h_1\}\right]\,.
\end{align}
We still follow an inductive scheme and assume that they hold for $p$ gravitons with $p\leqslant m-1$ gravitons. The $m=2$, $3$, and $4$ cases have been explicitly calculated in Section~\ref{sec:example}. For details, one can see the derivation of Eq.~\eqref{eq:2shuffle}, \eqref{eq:Bh2}, \eqref{eq:Pf2PTn-1}, \eqref{eq:psiT4} and \eqref{eq:psiU4}. It is sufficient to only work with $s=m-1$, since other values of $s$ can be obtained by a trivial exchange of indices. Like the calculation performed in Section~\ref{sec:4h}, we start with the expansion:
\begin{align}
    &\quad\;\frac{\sigma_{nh_{m-1}}}{\sigma_{nh_m}}\,\Pf(\Psi_{h_{m-1}\ldots h_1}^{n+m})\,\PT(12\ldots n)\nonumber\\
    &\cong\frac{\sigma_{nh_{m-1}}\sigma_{h_m1}}{\sigma_{n1}}\,(-1)^{m-1}\epsilon_{h_{m-1}}\cdot T\left[1,\{2\ldots n\}\shuffle\{h_{m-1}\},h_m|\{h_{m-2}\ldots h_1\}\right]\nonumber\\
    &\cong\frac{\sigma_{nh_{m-1}}\sigma_{h_m1}}{\sigma_{n1}}\,(-1)^{m-1}\epsilon_{h_{m-1}}\cdot T\left[1,\{2\ldots n-1\}\shuffle\{h_{m-1}\},n,h_m|\{h_{m-2}\ldots h_1\}\right]\nonumber\\
    &\quad-(-1)^{m+1}C_{h_{m-1}h_m}\,\Pf(\Psi_{h_{m-2}\ldots h_1}^{n+m})\,\PT(12\ldots n)\nonumber\\
    &\quad+(-1)^{m}\left(\frac{\sigma_{nh_{m-1}}\sigma_{h_{m-1}1}}{\sigma_{n1}}\right)\sum_{t=1}^{m-2}C_{h_{m-1}h_t}\Bigg[\frac{\sigma_{h_{m-1}h_t}}{\sigma_{h_{m-1}h_m}}\,\Pf(\Psi_{h_{m-2}\ldots h_1}^{n+m})\PT(12\ldots nh_{m-1})\nonumber\\
    &\qquad\qquad\qquad\qquad\qquad-\sigma_{h_mh_t}(-1)^{m-1}\epsilon_{h_t}\cdot T\left[1,\{2\ldots n\}\shuffle\{h_th_m\},h_{m-1}|\{h_{m-2}\ldots\slashed{h}_t\ldots h_1\}\right]\Bigg]\nonumber\\
    &\quad-\left(\frac{\sigma_{nh_{m-1}}\sigma_{h_{m-1}1}}{\sigma_{n1}}\right)\sum_{t=1}^{m-2}B_{h_{m-1}h_t}\sigma_{h_mh_t} k_{h_t}\cdot T\left[1,\{2\ldots n\}\shuffle\{h_th_m\},h_{m-1}|\{h_{m-2}\ldots\slashed{h}_t\ldots h_1\}\right]
\end{align}
Now our inductive assumption says:\footnote{The matrix $\psi(m,m-1,t|m-2\ldots\slashed{t}\ldots 1)$ is obtained from $\psi(m,m-1|m-2\ldots 1)$ by deleting the row and column intersected at $\pm C_{h_{m-1}h_t}$.}
\begin{align}
    &\quad\;\frac{1}{\sigma_{h_mh_t}}\left[\frac{\sigma_{h_{m-1}h_t}}{\sigma_{h_{m-1}h_m}}\Pf(\Psi_{h_{m-2}\ldots h_1}^{n+m})+\Pf\left[\psi(m,m-1,t|m-2\ldots\slashed{t}\ldots 1)\right]\right]\nonumber\\
    &\cong(-1)^{m-1}\epsilon_{h_t}\cdot T\left[1,\{2\ldots n\}\shuffle\{h_th_m\},h_{m-1}|\{h_{m-2}\ldots\slashed{h}_t\ldots h_1\}\right]\,,\nonumber\\
    &\quad\;\frac{1}{\sigma_{h_mh_t}}\Pf\left[\bar\psi(m,m-1,t|m-2\ldots\slashed{t}\ldots 1)\right]\nonumber\\
    &\cong(-1)^{m-1}k_{h_t}\cdot T\left[1,\{2\ldots n\}\shuffle\{h_th_m\},h_{m-1}|\{h_{m-2}\ldots\slashed{h}_t\ldots h_1\}\right]\,.
\end{align}
As a result, we get:
\begin{align}
    &\quad\;\frac{\sigma_{nh_{m-1}}}{\sigma_{nh_m}}\,\Pf(\Psi_{h_{m-1}\ldots h_1}^{n+m})\,\PT(12\ldots n)\nonumber\\
    &\cong\frac{\sigma_{nh_{m-1}}\sigma_{h_m1}}{\sigma_{n1}}\,(-1)^{m-1}\epsilon_{h_{m-1}}\cdot T\left[1,\{2\ldots n-1\}\shuffle\{h_{m-1}\},n,h_m|\{h_{m-2}\ldots h_1\}\right]\nonumber\\
    &\quad-(-1)^{m+1}C_{h_{m-1}h_m}\,\Pf(\Psi_{h_{m-2}\ldots h_1}^{n+m})\,\PT(12\ldots n)\nonumber\\
    &\quad+(-1)^{m+1}\sum_{t=1}^{m-2}C_{h_{m-1}h_t}\Pf\left[\psi(m,m-1,t|m-2\ldots\slashed{t}\ldots 1)\right]\PT(12\ldots n)\nonumber\\
    &\quad-(-1)^{m+1}\sum_{t=1}^{m-2}B_{h_{m-1}h_t}\Pf\left[\bar\psi(m,m-1,t|m-2\ldots\slashed{t}\ldots 1)\right]\PT(12\ldots n)\nonumber\\
    &\cong\sigma_{h_mh_{m-1}}\,(-1)^{m}\epsilon_{h_{m-1}}\cdot T\left[1,\{2\ldots n-1\}\shuffle\{h_{m-1}h_m\},n|\{h_{m-2}\ldots h_1\}\right]\nonumber\\
    &\quad -\Pf\left[\psi(m,m-1|m-2\ldots 1)\right]\PT(12\ldots n)\,,
\end{align}
which proves Eq.~\eqref{eq:Tm-1}. Then replacing $\epsilon_{h_{m-1}}$ by $k_{h_{m-1}}$, we also proves Eq.~\eqref{eq:Um-1}:
\begin{align*}
    &\quad\left(-1\right)^{m} k_{h_{m-1}}\cdot T\left[1,\{2\ldots n-1\}\shuffle\{h_{m-1}h_m\},n|\{h_{m-2}\ldots h_1\}\right]\\
    &\cong\frac{1}{\sigma_{h_mh_{m-1}}}\Pf\left[\bar\psi(m,m-1|m-2\ldots 1)\right]\PT(12\ldots n)\,.
\end{align*}
With the help of these two equations, we can immediately see that Eq.~\eqref{eq:hmstep1} gives exactly our statement:
\begin{equation}
    (-1)^{m}\,\Pf(\Psi_{\mathsf{H}})\,\PT(12\ldots n)\cong\epsilon_{h_m}\cdot T\left[1,\{2\ldots n-1\}\shuffle\{h_m\},n|\{h_{m-1}\ldots h_1\}\right]\,,
\end{equation}
 We have now completed the proof of our general recursive relation Eq.~\eqref{eq:hmresult}.

\section{Expansion in terms of pure YM amplitudes}\label{sec:GraphicRules}

If we keep using the recursive relation \eqref{eq:hmresult}, we can finally expand generic EYM tree amplitudes in terms of pure YM amplitudes. In this representation, the final expression does not enjoy the explicit permutation invariance of the gravitons. The reason is that in Eq.~\eqref{eq:hmresult}, we have defined a standard graviton order $\{h_mh_{m-1}\ldots h_1\}$. At each level of recursion, we always convert the first graviton remained in this standard order into a gluon according to the definition of $T$, which breaks the explicit permutation invariance.\footnote{The permutation invariant expression can be recovered after a straightforward, although tedious, symmetrization. We will not discuss it in this paper.} In this section, we give these expansions a graph theory picture, derived from our recursive expansion \eqref{eq:hmresult}. More details on this derivation is presented in Appendix~\ref{sec:derivation}

\subsection{Some expansion examples }

In this part, we will give the expansions of the EYM amplitudes with two and three gravitons. It serves two purposes: (1) establishing our convention; (2) providing explicit expressions to compare with known results as well as a better understanding of the graphic rules.

Let us start with the example of two gravitons, whose recursive construction is given in Eq.~\eqref{eq:2hamp}. Since the last term of Eq.~\eqref{eq:2hamp} contains only YM amplitudes. We only need to further expand the first term using Eq.~\eqref{eq:1hamp}:
\begin{align}
\label{eq:2hpureYM}
    &\quad \sum_{\shuffle}\left(\epsilon_{h_2}\cdot Y_{h_2}\right)A_{n+1,1}^{\text{EYM}}(1,\{2\ldots n-1\}\shuffle\{h_2\},n\,|\,h_1)\nonumber\\
    &=-\sum_{\shuffle}\left(\epsilon_{h_2}\cdot Y_{h_2}\right)\left(\epsilon_{h_1}\cdot X_{h_1}\right)A_{n+2}^{\text{YM}}(1,\big[\{2\ldots n-1\}\shuffle\{h_2\}\big]\shuffle\{h_1\},n)\,,
\end{align}
where $Y_{h_i}$ is defined as the sum of all original gluon momenta before the graviton $h_i$ in each ordering of the shuffle product. The important difference between $Y_{h_i}$ and $X_{h_i}$ is that $X_{h_i}$ is the sum of all momenta before $h_i$, including those gluons converted from gravitons. Now using the associativity of the shuffle product:
\begin{align}
    \big[\{2\ldots n-1\}\shuffle\{h_2\}\big]\shuffle\{h_1\}&=\{2\ldots n-1\}\shuffle\big[\{h_2\}\shuffle\{h_1\}\big]\nonumber\\
    &=\{2\ldots n-1\}\shuffle\{h_1h_2\}+\{2\ldots n-1\}\shuffle\{h_2h_1\}\,,
\end{align}
we can write the second line of Eq.~\eqref{eq:2hpureYM} as
\begin{align}
    &-\sum_{\shuffle}\left(\epsilon_{h_2}\cdot Y_{h_2}\right)\left(\epsilon_{h_1}\cdot Y_{h_1}\right)A_{n+2}^{\text{YM}}(1,\{2\ldots n-1\}\shuffle\{h_1h_2\},n)\nonumber\\
    &-\sum_{\shuffle}\left(\epsilon_{h_2}\cdot Y_{h_2}\right)\left(\epsilon_{h_1}\cdot Y_{h_1}+\epsilon_{h_1}\cdot k_{h_2}\right)A_{n+2}^{\text{YM}}(1,\{2\ldots n-1\}\shuffle\{h_2h_1\},n)\,.
\end{align}
Therefore, if we further expand Eq.~\eqref{eq:2hamp} down to the pure YM level, we get:
\begin{align}
\label{eq:2hnotsymm}
    (-1)A_{n,2}^{\text{EYM}}(12\ldots n\,|\,h_2h_1)&=\sum_{\shuffle}\left(\epsilon_{h_2}\cdot Y_{h_2}\right)\left(\epsilon_{h_1}\cdot Y_{h_1}\right)A_{n+2}^{\text{YM}}(1,\{2\ldots n-1\}\shuffle\{h_1h_2\},n)\nonumber\\
    &\quad+\sum_{\shuffle}\left(\epsilon_{h_2}\cdot Y_{h_2}\right)\left(\epsilon_{h_1}\cdot Y_{h_1}\right)A_{n+2}^{\text{YM}}(1,\{2\ldots n-1\}\shuffle\{h_2h_1\},n)\nonumber\\
    &\quad+\sum_{\shuffle}\left(\epsilon_{h_2}\cdot F_{h_1}\cdot Y_{h_1}\right)A_{n+2}^{\text{YM}}(1,\{2\ldots n-1\}\shuffle\{h_1h_2\},n)\nonumber\\
    &\quad+\sum_{\shuffle}\left(\epsilon_{h_1}\cdot k_{h_2}\right)\left(\epsilon_{h_2}\cdot Y_{h_2}\right)A_{n+2}^{\text{YM}}(1,\{2\ldots n-1\}\shuffle\{h_2h_1\},n)\,.
\end{align}
We want to emphasize that in the above expansion, the coefficients contain only the kinematic variables $\eps_{h_i}, k_{h_i}$ and $Y_{h_i}$. 

Similar manipulations can be performed for EYM amplitudes with three gravitons. By recursively using Eq.~\eqref{G2-manifest-1} and \eqref{EYM1gra} to Eq.~\eqref{threeGraRe}, one can write the final result as:\footnote{From now on, we are going to use the boldface $\pmb\sigma$ to stand for an ordered set, while $\sigma(i)$ denotes the $i$-th element in the set $\pmb\sigma$. Please do not confuse them with the solutions to the scattering equations.}
\begin{equation}
    A_{n,3}^{\text{EYM}}(12\ldots n\,|\,h_3h_2h_1)=\sum_{\pmb{\sigma}\in S_3}\sum_{\shuffle}C_{[321]}(\pmb{\sigma})A_{n+3}^{\text{YM}}(1,\{2\ldots n-1\}\shuffle\{h_{\sigma(1)}h_{\sigma(2)}h_{\sigma(3)}\},n)\,.
\end{equation}
For each graviton permutation $\pmb{\sigma}$, the coefficient $C_{[321]}(\pmb{\sigma})$ has the following expression:
\begingroup
\allowdisplaybreaks
\begin{align}
\label{eq:C123}
    C_{[321]}(123)&=\left(\epsilon_{h_3}\cdot Y_{h_3}\right)\left(\epsilon_{h_2}\cdot Y_{h_2}\right)\left(\epsilon_{h_1}\cdot Y_{h_1}\right)+\left(\epsilon_{h_3}\cdot F_{h_2}\cdot Y_{h_2}\right)\left(\epsilon_{h_1}\cdot Y_{h_1}\right)\nonumber\\*
    &\quad+\left(\epsilon_{h_3}\cdot F_{h_1}\cdot Y_{h_1}\right)\left(\epsilon_{h_2}\cdot Y_{h_2}\right)+\left(\epsilon_{h_2}\cdot F_{h_1}\cdot Y_{h_1}\right)\left(\epsilon_{h_3}\cdot Y_{h_3}\right)\nonumber\\*
    &\quad+\left(\epsilon_{h_3}\cdot F_{h_1}\cdot Y_{h_1}\right)\left(\epsilon_{h_2}\cdot k_{h_1}\right)+\left(\epsilon_{h_3}\cdot F_{h_2}\cdot F_{h_1}\cdot Y_{h_1}\right)\\
\label{eq:C213}
    C_{[321]}(213)&=\left(\epsilon_{h_3}\cdot Y_{h_3}\right)\left(\epsilon_{h_1}\cdot Y_{h_1}\right)\left(\epsilon_{h_2}\cdot Y_{h_2}\right)+\left(\epsilon_{h_3}\cdot F_{h_1}\cdot Y_{h_1}\right)\left(\epsilon_{h_2}\cdot Y_{h_2}\right)\nonumber\\*
    &\quad+\left(\epsilon_{h_3}\cdot F_{h_2}\cdot Y_{h_2}\right)\left(\epsilon_{h_1}\cdot Y_{h_1}\right)+\left(\epsilon_{h_1}\cdot k_{h_2}\right)\left(\epsilon_{h_2}\cdot Y_{h_2}\right)\left(\epsilon_{h_3}\cdot Y_{h_3}\right)\nonumber\\*
    &\quad+\left(\epsilon_{h_3}\cdot F_{h_2}\cdot Y_{h_2}\right)\left(\epsilon_{h_1}\cdot k_{h_2}\right)+\left(\epsilon_{h_3}\cdot F_{h_1}\cdot F_{h_2}\cdot Y_{h_2}\right)\\
    C_{[321]}(321)&=\left(\epsilon_{h_1}\cdot Y_{h_1}\right)\left(\epsilon_{h_2}\cdot Y_{h_2}\right)\left(\epsilon_{h_3}\cdot Y_{h_3}\right)+\left(\epsilon_{h_1}\cdot k_{h_2}\right)\left(\epsilon_{h_2}\cdot Y_{h_2}\right)\left(\epsilon_{h_3}\cdot Y_{h_3}\right)\nonumber\\*
    &\quad+\left(\epsilon_{h_1}\cdot k_{h_3}\right)\left(\epsilon_{h_3}\cdot Y_{h_3}\right)\left(\epsilon_{h_2}\cdot Y_{h_2}\right)+\left(\epsilon_{h_2}\cdot k_{h_3}\right)\left(\epsilon_{h_3}\cdot Y_{h_3}\right)\left(\epsilon_{h_1}\cdot Y_{h_1}\right)\nonumber\\*
    &\quad+\left(\epsilon_{h_1}\cdot k_{h_3}\right)\left(\epsilon_{h_3}\cdot Y_{h_3}\right)\left(\epsilon_{h_2}\cdot k_{h_3}\right)+\left(\epsilon_{h_1}\cdot k_{h_2}\right)\left(\epsilon_{h_2}\cdot k_{h_3}\right)\left(\epsilon_{h_3}\cdot Y_{h_3}\right)\\
    C_{[321]}(132)&=\left(\epsilon_{h_2}\cdot Y_{h_2}\right)\left(\epsilon_{h_3}\cdot Y_{h_3}\right)\left(\epsilon_{h_1}\cdot Y_{h_1}\right)+\left(\epsilon_{h_2}\cdot k_{h_3}\right)\left(\epsilon_{h_3}\cdot Y_{h_3}\right)\left(\epsilon_{h_1}\cdot Y_{h_1}\right)\nonumber\\*
    &\quad+\left(\epsilon_{h_2}\cdot F_{h_1}\cdot Y_{h_1}\right)\left(\epsilon_{h_3}\cdot Y_{h_3}\right)+\left(\epsilon_{h_3}\cdot F_{h_1}\cdot Y_{h_1}\right)\left(\epsilon_{h_2}\cdot Y_{h_2}\right)\nonumber\\*
    &\quad+\left(\epsilon_{h_2}\cdot k_{h_1}\right)\left(\epsilon_{h_3}\cdot F_{h_1}\cdot Y_{h_1}\right)+\left(\epsilon_{h_2}\cdot k_{h_3}\right)\left(\epsilon_{h_3}\cdot F_{h_1}\cdot Y_{h_1}\right)\\
    C_{[321]}(231)&=\left(\epsilon_{h_1}\cdot Y_{h_1}\right)\left(\epsilon_{h_3}\cdot Y_{h_3}\right)\left(\epsilon_{h_2}\cdot Y_{h_2}\right)+\left(\epsilon_{h_1}\cdot k_{h_3}\right)\left(\epsilon_{h_3}\cdot Y_{h_3}\right)\left(\epsilon_{h_2}\cdot Y_{h_2}\right)\nonumber\\
    &\quad+\left(\epsilon_{h_1}\cdot k_{h_2}\right)\left(\epsilon_{h_2}\cdot Y_{h_2}\right)\left(\epsilon_{h_3}\cdot Y_{h_3}\right)+\left(\epsilon_{h_3}\cdot F_{h_2}\cdot Y_{h_2}\right)\left(\epsilon_{h_1}\cdot Y_{h_1}\right)\nonumber\\*
    &\quad+\left(\epsilon_{h_1}\cdot k_{h_2}\right)\left(\epsilon_{h_3}\cdot F_{h_2}\cdot Y_{h_2}\right)+\left(\epsilon_{h_1}\cdot k_{h_3}\right)\left(\epsilon_{h_3}\cdot F_{h_2}\cdot Y_{h_2}\right)\\
\label{eq:C312}
    C_{[321]}(312)&=\left(\epsilon_{h_2}\cdot Y_{h_2}\right)\left(\epsilon_{h_1}\cdot Y_{h_1}\right)\left(\epsilon_{h_3}\cdot Y_{h_3}\right)+\left(\epsilon_{h_2}\cdot F_{h_1}\cdot Y_{h_1}\right)\left(\epsilon_{h_3}\cdot Y_{h_3}\right)\nonumber\\*
    &\quad+\left(\epsilon_{h_2}\cdot k_{h_3}\right)\left(\epsilon_{h_3}\cdot Y_{h_3}\right)\left(\epsilon_{h_1}\cdot Y_{h_1}\right)+\left(\epsilon_{h_1}\cdot k_{h_3}\right)\left(\epsilon_{h_3}\cdot Y_{h_3}\right)\left(\epsilon_{h_2}\cdot Y_{h_2}\right)\nonumber\\*
    &\quad+\left(\epsilon_{h_2}\cdot k_{h_3}\right)\left(\epsilon_{h_3}\cdot Y_{h_3}\right)\left(\epsilon_{h_1}\cdot k_{h_3}\right)+\left(\epsilon_{h_2}\cdot F_{h_1}\cdot k_{h_3}\right)\left(\epsilon_{h_3}\cdot Y_{h_3}\right)\,.
\end{align}
\endgroup
The subscript $[321]$ stands for the standard reference order of converting the gravitons into gluons, whose role will be discussed later. Note that these coefficients under the standard order $[321]$ do not enjoy the explicit permutation invariance, namely,
\begin{equation*}
    C_{[321]}(213)\neq \left.C_{[321]}(123)\right|_{h_1\leftrightarrow h_2}\qquad C_{[321]}(321)\neq \left.C_{[321]}(123)\right|_{h_1\leftrightarrow h_3}\,,
\end{equation*}
for example. This is because the expansion basis is KK, which has some redundancy, such that a certain gauge choice is allowed.

It is worth to compare our $C_{[321]}(\pmb{\sigma})$ with the results given in a previous paper~\cite{Fu:2017uzt}, where the kinematic variable $Z$ has been used and the coefficients contain different number of terms. Here, although $C_{[321]}(\pmb{\sigma})$ are not symmetric under relabeling, the total number of terms contained in each $C_{[321]}(\pmb{\sigma})$ is the same, namely, $6=3!$ terms for this three-graviton example. In fact, with some further observations, we find the following features:
\begin{itemize}

\item In terms of only the $\eps$, $F$, and $Y$ variables, each coefficient is given by a sum of six terms. Then each of these six terms is a product of a few scalar factors, enclosed in parentheses.

\item Now we count the number of elements in each parenthesis. For examples, $ \left(\epsilon_{h_3}\cdot F_{h_2}\cdot F_{h_1}\cdot Y_{h_1}\right)$ has $4$ elements, $\left(\epsilon_{h_2}\cdot F_{h_1}\cdot k_{h_3}\right)$ has $3$ and $\left(\epsilon_{h_3}\cdot Y_{h_3}\right)$ has $2$. Then we find that for each term that appears in any $C_{[321]}(\pmb{\sigma})$, we have:
\bea \sum_{\rm all~factors} (n_i-1)=3\,.\eea
\end{itemize}
If we represent each graviton by a vertex, and supplement them with a common root, we can then associate each factor with a path starting at a graviton vertex labeled by $\epsilon$ and ending at either a graviton vertex (labeled by $k$) or the root (labeled by $Y)$, while $n_i$ is just the number of vertices along the path. One can immediately observe that each term of $C_{[321]}(\pmb{\sigma})$ provides a tree structure. The sum $\sum_{\rm all~factors} (n_i-1)$ gives nothing but the total number of gravitons. This observation motivates us to construct the following graph theory rules based on spanning trees.


%

\subsection{Graphic rules for coefficients of pure YM expansion}

In this part, we propose a set of graphic rules for a direct reading of the pure YM expansion coefficients, based on spanning trees. The recursive relation \eqref{eq:hmresult} indicates that for a generic $m$-graviton EYM amplitude, we can always expand it in terms of pure YM amplitudes in the KK basis, with the following form:\footnote{As kept implicit here, $C_{\pmb{\rho}}(\pmb{\sigma})$ also depends on the shuffle product, since it contains $Y_{h_i}$, the sum of gluon momenta before $h_i$ in $\{2\ldots n-1\}\shuffle\pmb{\sigma}$.}
\begin{equation}
\label{eq:hmexpansion}
    A_{n,m}^{\text{EYM}}(12\ldots n\,|\,h_1h_2\ldots h_m)=\sum_{\pmb{\sigma}\in S_m}\sum_{\shuffle}C_{\pmb{\rho}}(\pmb{\sigma})A_{n+m}^{\text{YM}}(1,\{2\ldots n-1\}\shuffle\pmb{\sigma},n)\,,
\end{equation}
where $\pmb{\sigma}$ is a permutation of the $m$ gravitons $\{h_1\ldots h_m\}$ and $\pmb{\rho}$ is a reference order. The coefficients $C_{\pmb{\rho}}(\pmb{\sigma})$ can be evaluated from the spanning trees of $m+1$ vertices. Among these vertices, $m$ of them correspond to the gravitons $\{h_1,h_2\ldots h_m\}$, and a single vertex $\mathsf{g}$ as the root, representing the gluon set $\{2\ldots n-1\}$. The coefficients $C_{\pmb{\rho}}(\pmb{\sigma})$ can be read out by the following algorithm:

~\\{\bf Step 1: constructing the trees contributing to the order $\pmb\sigma$.}   We construct the \emph{increasing trees} with respect to the order $\pmb\sigma$, i.e., $\mathsf{g}\prec h_{\sigma(1)}\prec h_{\sigma(2)}\prec\ldots\prec h_{\sigma(m)}$. Such trees can be obtained by arranging vertices $h_{\sigma(1)}, h_{\sigma(2)},\ldots, h_{\sigma(m)}$ from the left to right and then drawing all possible tree diagrams after avoiding the following situations: \emph{if two vertices $A\prec B$ are connected by a path originated from the root, then $B$ can not be the one closer to the root}. {We emphasize that the construction of the trees only depends on $\pmb{\sigma}$, not the reference order $\pmb{\rho}$.}

~\\
For example, when there are three gravitons, we have the following trees contributing to $\pmb{\sigma}=\{123\}$:
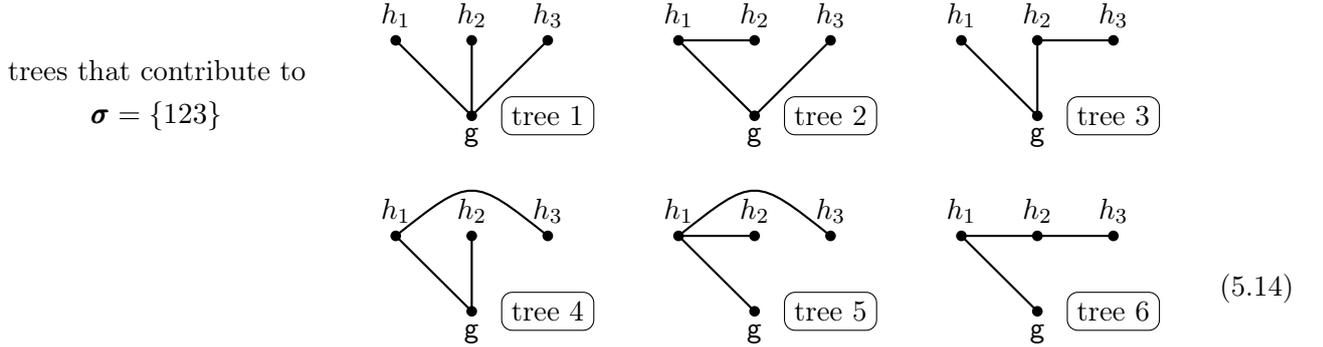
\begin{align}
  \begin{array}{c}
    \text{trees that contribute to} \\
    \pmb{\sigma}=\{123\} \\
    \end{array}& &\adjustbox{raise=-0.75cm}{\begin{tikzpicture}
    \coordinate (g) at (0,0);
    \fill (g) circle (2pt) node[below=1pt]{$\mathsf{g}$};
    \coordinate (h1) at (-1,1);
    \fill (h1) circle (2pt) node[above=1pt]{$h_1$};
    \coordinate (h2) at (0,1);
    \fill (h2) circle (2pt) node[above=1pt]{$h_2$};
    \coordinate (h3) at (1,1);
    \fill (h3) circle (2pt) node[above=1pt]{$h_3$};
    \draw [thick] (g) -- (h1) (g) -- (h2) (g) -- (h3);
    \node at (1,0) [draw,rectangle,rounded corners] {tree 1};
  \end{tikzpicture}}
  & &\adjustbox{raise=-0.75cm}{\begin{tikzpicture}
    \coordinate (g) at (0,0);
    \fill (g) circle (2pt) node[below=1pt]{$\mathsf{g}$};
    \coordinate (h1) at (-1,1);
    \fill (h1) circle (2pt) node[above=1pt]{$h_1$};
    \coordinate (h2) at (0,1);
    \fill (h2) circle (2pt) node[above=1pt]{$h_2$};
    \coordinate (h3) at (1,1);
    \fill (h3) circle (2pt) node[above=1pt]{$h_3$};
    \draw [thick] (g) -- (h1) -- (h2) (g) -- (h3);
    \node at (1,0) [draw,rectangle,rounded corners] {tree 2};
  \end{tikzpicture}}
  & &\adjustbox{raise=-0.75cm}{\begin{tikzpicture}
    \coordinate (g) at (0,0);
    \fill (g) circle (2pt) node[below=1pt]{$\mathsf{g}$};
    \coordinate (h1) at (-1,1);
    \fill (h1) circle (2pt) node[above=1pt]{$h_1$};
    \coordinate (h2) at (0,1);
    \fill (h2) circle (2pt) node[above=1pt]{$h_2$};
    \coordinate (h3) at (1,1);
    \fill (h3) circle (2pt) node[above=1pt]{$h_3$};
    \draw [thick] (g) -- (h1) (g) -- (h2) -- (h3);
    \node at (1,0) [draw,rectangle,rounded corners] {tree 3};
    \end{tikzpicture}}\nonumber\\
    & &\adjustbox{raise=-0.75cm}{\begin{tikzpicture}
    \coordinate (g) at (0,0);
    \fill (g) circle (2pt) node[below=1pt]{$\mathsf{g}$};
    \coordinate (h1) at (-1,1);
    \fill (h1) circle (2pt) node[above=1pt]{$h_1$};
    \coordinate (h2) at (0,1);
    \fill (h2) circle (2pt) node[above=1pt]{$h_2$};
    \coordinate (h3) at (1,1);
    \fill (h3) circle (2pt) node[above=1pt]{$h_3$};
    \draw [thick] (g) -- (h1) .. controls (0,1.8) and (0,1.8) .. (h3) (g) -- (h2);
    \node at (1,0) [draw,rectangle,rounded corners] {tree 4};
  \end{tikzpicture}}
  & &\adjustbox{raise=-0.75cm}{\begin{tikzpicture}
    \coordinate (g) at (0,0);
    \fill (g) circle (2pt) node[below=1pt]{$\mathsf{g}$};
    \coordinate (h1) at (-1,1);
    \fill (h1) circle (2pt) node[above=1pt]{$h_1$};
    \coordinate (h2) at (0,1);
    \fill (h2) circle (2pt) node[above=1pt]{$h_2$};
    \coordinate (h3) at (1,1);
    \fill (h3) circle (2pt) node[above=1pt]{$h_3$};
    \draw [thick] (g) -- (h1) .. controls (0,1.8) and (0,1.8) .. (h3) (h1) -- (h2);
    \node at (1,0) [draw,rectangle,rounded corners] {tree 5};
  \end{tikzpicture}}
  & &\adjustbox{raise=-0.75cm}{\begin{tikzpicture}
    \coordinate (g) at (0,0);
    \fill (g) circle (2pt) node[below=1pt]{$\mathsf{g}$};
    \coordinate (h1) at (-1,1);
    \fill (h1) circle (2pt) node[above=1pt]{$h_1$};
    \coordinate (h2) at (0,1);
    \fill (h2) circle (2pt) node[above=1pt]{$h_2$};
    \coordinate (h3) at (1,1);
    \fill (h3) circle (2pt) node[above=1pt]{$h_3$};
    \draw [thick] (g) -- (h1) -- (h2) -- (h3);
    \node at (1,0) [draw,rectangle,rounded corners] {tree 6};
    \end{tikzpicture}}
\label{C123-tree}\end{align}
In these trees, all vertices in a path starting from the root and ending at a leaf respect the order $\pmb{\sigma}$. There are exactly $m!$ such trees. This can be easily seen by a recursive construction. For example, from each tree shown in Eq.~\eqref{C123-tree}, we can construct exactly four increasing trees for $\{1234\}$ by either connecting $h_4$ to any $h_{i}$, with $i=1,2,3$, or connecting $h_4$ directly to the root. More explicitly, the first tree gives:
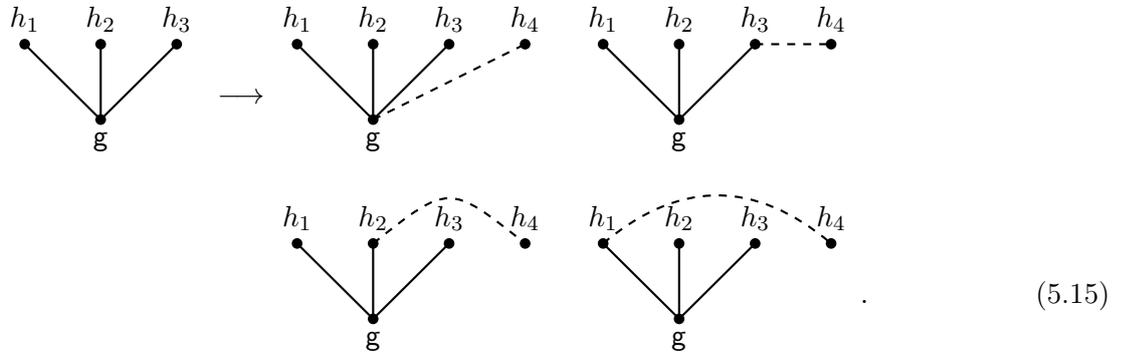
\begin{align}
    \adjustbox{raise=-0.75cm}{\begin{tikzpicture}
    \coordinate (g) at (0,0);
    \fill (g) circle (2pt) node[below=1pt]{$\mathsf{g}$};
    \coordinate (h1) at (-1,1);
    \fill (h1) circle (2pt) node[above=1pt]{$h_1$};
    \coordinate (h2) at (0,1);
    \fill (h2) circle (2pt) node[above=1pt]{$h_2$};
    \coordinate (h3) at (1,1);
    \fill (h3) circle (2pt) node[above=1pt]{$h_3$};
    \draw [thick] (g) -- (h1) (g) -- (h2) (g) -- (h3);
    \end{tikzpicture}}\;\longrightarrow\;&\adjustbox{raise=-0.75cm}{\begin{tikzpicture}
    \coordinate (g) at (0,0);
    \fill (g) circle (2pt) node[below=1pt]{$\mathsf{g}$};
    \coordinate (h1) at (-1,1);
    \fill (h1) circle (2pt) node[above=1pt]{$h_1$};
    \coordinate (h2) at (0,1);
    \fill (h2) circle (2pt) node[above=1pt]{$h_2$};
    \coordinate (h3) at (1,1);
    \fill (h3) circle (2pt) node[above=1pt]{$h_3$};
    \coordinate (h4) at (2,1);
    \fill (h4) circle (2pt) node[above=1pt]{$h_4$};
    \draw [thick] (g) -- (h1) (g) -- (h2) (g) -- (h3);
    \draw [thick,dashed] (g) -- (h4);
    \end{tikzpicture}}\quad\adjustbox{raise=-0.75cm}{\begin{tikzpicture}
    \coordinate (g) at (0,0);
    \fill (g) circle (2pt) node[below=1pt]{$\mathsf{g}$};
    \coordinate (h1) at (-1,1);
    \fill (h1) circle (2pt) node[above=1pt]{$h_1$};
    \coordinate (h2) at (0,1);
    \fill (h2) circle (2pt) node[above=1pt]{$h_2$};
    \coordinate (h3) at (1,1);
    \fill (h3) circle (2pt) node[above=1pt]{$h_3$};
    \coordinate (h4) at (2,1);
    \fill (h4) circle (2pt) node[above=1pt]{$h_4$};
    \draw [thick] (g) -- (h1) (g) -- (h2) (g) -- (h3);
    \draw [thick,dashed] (h3) -- (h4);
    \end{tikzpicture}}\nonumber\\
    &\adjustbox{raise=-0.75cm}{\begin{tikzpicture}
    \coordinate (g) at (0,0);
    \fill (g) circle (2pt) node[below=1pt]{$\mathsf{g}$};
    \coordinate (h1) at (-1,1);
    \fill (h1) circle (2pt) node[above=1pt]{$h_1$};
    \coordinate (h2) at (0,1);
    \fill (h2) circle (2pt) node[above=1pt]{$h_2$};
    \coordinate (h3) at (1,1);
    \fill (h3) circle (2pt) node[above=1pt]{$h_3$};
    \coordinate (h4) at (2,1);
    \fill (h4) circle (2pt) node[above=1pt]{$h_4$};
    \draw [thick] (g) -- (h1) (g) -- (h2) (g) -- (h3);
    \draw [thick,dashed] (h2) .. controls (1,1.8) and (1,1.8) .. (h4);
    \end{tikzpicture}}\quad\adjustbox{raise=-0.75cm}{\begin{tikzpicture}
    \coordinate (g) at (0,0);
    \fill (g) circle (2pt) node[below=1pt]{$\mathsf{g}$};
    \coordinate (h1) at (-1,1);
    \fill (h1) circle (2pt) node[above=1pt]{$h_1$};
    \coordinate (h2) at (0,1);
    \fill (h2) circle (2pt) node[above=1pt]{$h_2$};
    \coordinate (h3) at (1,1);
    \fill (h3) circle (2pt) node[above=1pt]{$h_3$};
    \coordinate (h4) at (2,1);
    \fill (h4) circle (2pt) node[above=1pt]{$h_4$};
    \draw [thick] (g) -- (h1) (g) -- (h2) (g) -- (h3);
    \draw [thick,dashed] (h1) .. controls (0,1.85) and (1,1.85) .. (h4);
    \end{tikzpicture}}\,.
\end{align}

~\\{ \bf Step2: Reading out the expression for a tree.} Having constructed the increasing trees, we assign an expression to each tree. At this moment, the choice of ordering ${\pmb \rho}$ will play an important role. We will accomplish it by
following steps:
\begin{itemize}

\item First, we draw ordered paths. We will start from  $\rho(1)$, the first graviton in list ${\pmb \rho}$, and then draw a path from the root $\mathsf{g}$ to $\rho(1)$. If there are $\ell$ vertices along this path, we will set $\phi_1=\rho(1)$, and the subsequent vertices $\phi_2$, $\phi_3$, etc, until the root $\phi_\ell=\mathsf{g}$.
We will denote such a path by ${\cal P}[1]=\{\rho(1),\phi_2,...,\phi_{\ell-1},\mathsf{g}\}$. An illustration of such
path ${\cal P}[1]$ is given as the following:
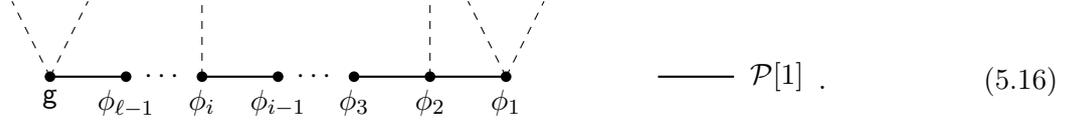
\begin{equation}
    \label{eq:path}
        \adjustbox{raise=-0.5cm}{\begin{tikzpicture}
            \coordinate (g) at (0,0);
            \fill (g) circle (2pt) node[below=1pt]{$\mathsf{g}$};
            \coordinate (a) at (1,0);
            \fill (a) circle (2pt) node[below=1pt]{$\phi_{\ell-1}$};
            \draw [thick] (g) -- (a);
            \node at (1.5,0) {$\cdots$};
            \coordinate (b) at (2,0);
            \fill (b) circle (2pt) node[below=1pt]{$\phi_i$};
            \coordinate (c) at (3,0);
            \fill (c) circle (2pt) node[below=1pt]{$\phi_{i-1}$};
            \draw [thick] (b) -- (c);
            \node at (3.5,0) {$\cdots$};
            \coordinate (d) at (4,0);
            \fill (d) circle (2pt) node[below=1pt]{$\phi_3$};
            \coordinate (e) at (5,0);
            \fill (e) circle (2pt) node[below=1pt]{$\phi_2$};
            \coordinate (f) at (6,0);
            \fill (f) circle (2pt) node[below=1pt]{$\phi_1$};
            \draw [thick] (d) -- (f);
            \draw [dashed] (e) -- ++(0,1);
            \draw [dashed] (b) -- ++(0,1);
            \draw [dashed] (g) -- ++(0.5,1) (g) -- ++(-0.5,1);
            \draw [dashed] (f) -- ++(0.5,1) (f) -- ++(-0.5,1);
            \draw [thick,black] (8,0) -- ++(1,0) node[right=2pt]{${\cal P}[1]$};
        \end{tikzpicture}}\,.
    \end{equation}
It is worth to emphasize that $\rho(1)$ may or may not be a leaf. 

\item As the first path is done, we construct the second path from the remaining vertices:
\begin{equation}
        {\pmb \rho}_1\equiv {\pmb \rho}\backslash\{\phi_1\ldots \phi_{\ell-1}, \mathsf{g}\}\,,
    \end{equation}

Now we construct ${\cal P}[2]$ by considering the path from the root $ \mathsf{g}$ to the first element of $ {\pmb \rho}_{1}$, denoted as $\rho_1(1)$. Let us denote this tentative path as ${\cal T}_2$. Starting from the root, the path ${\cal T}_2$ will in general coincide with other previous constructed paths until the vertex $V_{2}$, the last vertex (counted from the root) along ${\cal T}_2$ that belongs to previous constructed paths. The crossroad vertex $V_2$ can either be a graviton or just the root. Now the part of ${\cal T}_2$ from $V_2$ and beyond, until $\rho_1(1)$, is our ${\cal P}[2]$. It has the form ${\cal P}[2]=\{\W\phi_1,\W\phi_2,...,\W\phi_{t},V_{2}\}$, where $\W\phi_i\in {\pmb \rho_1}$ but $V_{2}\not\in {\pmb \rho_1}$. This process is illustrated as:
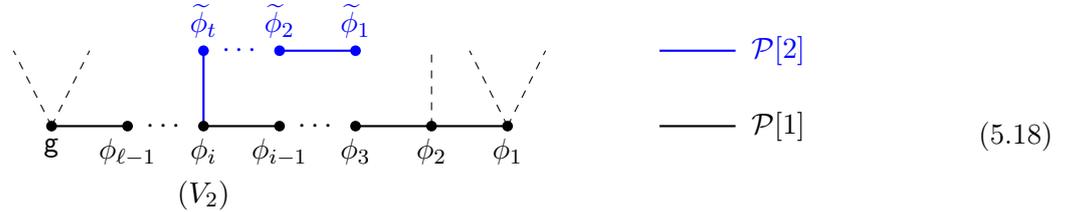
\begin{equation}
    \label{eq:path2}
        \adjustbox{raise=-1cm}{\begin{tikzpicture}
            \coordinate (g) at (0,0);
            \fill (g) circle (2pt) node[below=1pt]{$\mathsf{g}$};
            \coordinate (a) at (1,0);
            \fill (a) circle (2pt) node[below=1pt]{$\phi_{\ell-1}$};
            \draw [thick] (g) -- (a);
            \node at (1.5,0) {$\cdots$};
            \coordinate (b) at (2,0);
            \draw [thick,blue] (b) -- ++(0,1);
            \fill (b) circle (2pt) node[below=1pt,align=center]{$\phi_i$ \\ $(V_2)$};
            \coordinate (c) at (3,0);
            \fill (c) circle (2pt) node[below=1pt]{$\phi_{i-1}$};
            \draw [thick] (b) -- (c);
            \node at (3.5,0) {$\cdots$};
            \coordinate (d) at (4,0);
            \fill (d) circle (2pt) node[below=1pt]{$\phi_3$};
            \coordinate (e) at (5,0);
            \fill (e) circle (2pt) node[below=1pt]{$\phi_2$};
            \coordinate (f) at (6,0);
            \fill (f) circle (2pt) node[below=1pt]{$\phi_1$};
            \coordinate (h) at (2,1);
            \fill [blue] (h) circle (2pt) node[above=1pt]{$\W{\phi}_t$};
            \node at (2.5,1) [text=blue]{$\cdots$};
            \coordinate (i) at (3,1);
            \fill [blue] (i) circle (2pt) node[above=1pt]{$\W{\phi}_2$};
            \coordinate (j) at (4,1);
            \fill [blue] (j) circle (2pt) node[above=1pt]{$\W{\phi}_1$};
            \draw [thick,blue] (i) -- (j);
            \draw [thick] (d) -- (f);
            \draw [dashed] (e) -- ++(0,1);
            \draw [dashed] (g) -- ++(0.5,1) (g) -- ++(-0.5,1);
            \draw [dashed] (f) -- ++(0.5,1) (f) -- ++(-0.5,1);
            \draw [thick,black] (8,0) -- ++(1,0) node[right=2pt]{${\cal P}[1]$};
            \draw [thick,blue] (8,1) -- ++(1,0) node[right=2pt]{${\cal P}[2]$};
        \end{tikzpicture}}
    \end{equation}
\item Repeat this procedure until we have exhausted all the graviton vertices.
\item  Now we assign each path a factor. The first element in the path is replaced by $\eps$, the middle element by $F$. For the last element, if it is the root, then we replace it by by $Y_{t}$ if it is the root $\mathsf{g}$ (the $t$ is the next-to-last element); if it is a graviton vertex $a$, we replace it by $k_{a}$. Finally, multiplying all these factors together, we get the term contributed by the tree under our consideration.
\item {We emphasize that given a tree, the evaluation only depends on the reference order $\pmb{\rho}$. Actually, a tree can contribute to more than one $\pmb{\sigma}$ in general, which will be further clarified later in this section.}
\end{itemize}
To demonstrate how these rules work, we again take ${\pmb \rho}=\{321\}$ and consider the six trees given in \eref{C123-tree} as an example. {These trees all contribute to $\pmb{\sigma}=\{123\}$.} It is easy to see that we have
\bea {\rm tree~1}: & ~~ & {\cal P}[1]= \{h_3,\mathsf{g}\};~~~~{\cal P}[2]= \{h_2,\mathsf{g}\};~~~~{\cal P}[3]= \{h_1,\mathsf{g}\};~~~~\nn
{\rm tree~2}: & ~~ & {\cal P}[1]= \{h_3,\mathsf{g}\};~~~~{\cal P}[2]= \{h_2,h_1,\mathsf{g}\};~~~~\nn
{\rm tree~3}: & ~~ & {\cal P}[1]= \{h_3,h_2,\mathsf{g}\};~~~~{\cal P}[2]= \{h_1,\mathsf{g}\}; \nn
{\rm tree~4}: & ~~ & {\cal P}[1]= \{h_3,h_1,\mathsf{g}\};~~~~{\cal P}[2]= \{h_2,\mathsf{g}\}; \nn
{\rm tree~5}: & ~~ & {\cal P}[1]= \{h_3,h_1,\mathsf{g}\};~~~~{\cal P}[2]= \{h_2,h_1\};\nn
{\rm tree~6}: & ~~ & {\cal P}[1]= \{h_3, h_2, h_1,\mathsf{g}\};~~~\Label{C123-pathes}\eea
Using our rules, we can write down six coefficients given in Eq.~\eqref{C123-pathes}
\bea {\rm tree~1}: & ~~ &  (\eps_{h_3}\cdot Y_{h_3})\times (\eps_{h_2}\cdot Y_{h_2})\times(\eps_{h_1}\cdot Y_{h_1});~~~~\nn
{\rm tree~2}: & ~~ &  (\eps_{h_3}\cdot Y_{h_3})\times (\eps_{h_2}\cdot F_{h_1}\cdot Y_{h_1});\nn
{\rm tree~3}: & ~~ & (\eps_{h_3}\cdot F_{h_2}\cdot Y_{h_2})\times(\eps_{h_1}\cdot Y_{h_1}); \nn
{\rm tree~4}: & ~~ & (\eps_{h_3}\cdot F_{h_1}\cdot Y_{h_1})\times(\eps_{h_2}\cdot Y_{h_1}); \nn
{\rm tree~5}: & ~~ & (\eps_{h_3}\cdot F_{h_1}\cdot Y_{h_1})\times (\eps_{h_2}\cdot k_{h_1});\nn
{\rm tree~6}: & ~~ &  (\eps_{h_3}\cdot F_{h_2}\cdot F_{h_1}\cdot Y_{h_1});~~~\Label{C123-pathes-factor}\eea
One can check that they exactly reproduce the $C_{[321]}(123)$ given in Eq.~\eqref{eq:C123}. With the reference order $\pmb{\rho}=\{321\}$, another example is
\begingroup
\allowdisplaybreaks
\begin{align}
 \label{321-tree}
 \adjustbox{raise=-0.75cm}{\begin{tikzpicture}
    \coordinate (g) at (0,0);
    \fill (g) circle (2pt) node[below=1pt]{$\mathsf{g}$};
    \coordinate (h1) at (-1,1);
    \fill (h1) circle (2pt) node[above=1pt]{$h_1$};
    \coordinate (h2) at (0,1);
    \fill (h2) circle (2pt) node[above=1pt]{$h_2$};
    \coordinate (h3) at (1,1);
    \fill (h3) circle (2pt) node[above=1pt]{$h_3$};
    \draw [thick] (g) -- (h3) -- (h2) -- (h1);
    \end{tikzpicture}}&=\left(\epsilon_{h_3}\cdot Y_{h_3}\right)\left(\epsilon_{h_2}\cdot k_{h_3}\right)\left(\epsilon_{h_1}\cdot k_{h_2}\right)\,.
\end{align}
{This tree contributes to $\pmb{\sigma}=\{321\}$ only.} As some less trivial examples, we consider several typical trees that appear in the evaluation of the coefficients for four-graviton EYM amplitudes. With the reference order $\pmb{\rho}=\{4321\}$, we have:
\begin{align}
    \adjustbox{raise=-0.75cm}{\begin{tikzpicture}
    \coordinate (g) at (0,0);
    \fill (g) circle (2pt) node[below=1pt]{$\mathsf{g}$};
    \coordinate (h1) at (-1.5,1);
    \fill (h1) circle (2pt) node[above=1pt]{$h_1$};
    \coordinate (h2) at (-0.5,1);
    \fill (h2) circle (2pt) node[above=1pt]{$h_2$};
    \coordinate (h3) at (0.5,1);
    \fill (h3) circle (2pt) node[above=1pt]{$h_3$};
    \coordinate (h4) at (1.5,1);
    \fill (h4) circle (2pt) node[above=1pt]{$h_4$};
    \draw [thick] (g) -- (h3) -- (h2) (h3) -- (h4) (h1) .. controls (0,2) and (0,2) .. (h4);
    \end{tikzpicture}}&=\left(\epsilon_{h_4}\cdot F_{h_3}\cdot Y_{h_3}\right)\left(\epsilon_{h_2}\cdot k_{h_3}\right)\left(\epsilon_{h_1}\cdot k_{h_4}\right)\nonumber\\
    \adjustbox{raise=-0.75cm}{\begin{tikzpicture}
    \coordinate (g) at (0,0);
    \fill (g) circle (2pt) node[below=1pt]{$\mathsf{g}$};
    \coordinate (h1) at (-1.5,1);
    \fill (h1) circle (2pt) node[above=1pt]{$h_1$};
    \coordinate (h2) at (-0.5,1);
    \fill (h2) circle (2pt) node[above=1pt]{$h_2$};
    \coordinate (h3) at (0.5,1);
    \fill (h3) circle (2pt) node[above=1pt]{$h_3$};
    \coordinate (h4) at (1.5,1);
    \fill (h4) circle (2pt) node[above=1pt]{$h_4$};
    \draw [thick] (g) -- (h4) -- (h3) -- (h2) (h1) .. controls (0,2) and (0,2) .. (h4);
    \end{tikzpicture}}&=\left(\epsilon_{h_4}\cdot Y_{h_4}\right)\left(\epsilon_{h_3}\cdot k_{h_4}\right)\left(\epsilon_{h_2}\cdot k_{h_3}\right)\left(\epsilon_{h_1}\cdot k_{h_4}\right)\nonumber\\
    \adjustbox{raise=-0.75cm}{\begin{tikzpicture}
    \coordinate (g) at (0,0);
    \fill (g) circle (2pt) node[below=1pt]{$\mathsf{g}$};
    \coordinate (h1) at (-1.5,1);
    \fill (h1) circle (2pt) node[above=1pt]{$h_1$};
    \coordinate (h2) at (-0.5,1);
    \fill (h2) circle (2pt) node[above=1pt]{$h_2$};
    \coordinate (h3) at (0.5,1);
    \fill (h3) circle (2pt) node[above=1pt]{$h_3$};
    \coordinate (h4) at (1.5,1);
    \fill (h4) circle (2pt) node[above=1pt]{$h_4$};
    \draw [thick] (g) -- (h3) -- (h4) (g) -- (h2) -- (h1);
    \end{tikzpicture}}&=\left(\epsilon_{h_4}\cdot F_{h_3}\cdot Y_{h_3}\right)\left(\epsilon_{h_2}\cdot Y_{h_2}\right)\left(\epsilon_{h_1}\cdot k_{h_2}\right)\nonumber\\
        \adjustbox{raise=-0.75cm}{\begin{tikzpicture}
    \coordinate (g) at (0,0);
    \fill (g) circle (2pt) node[below=1pt]{$\mathsf{g}$};
    \coordinate (h1) at (-1.5,1);
    \fill (h1) circle (2pt) node[above=1pt]{$h_1$};
    \coordinate (h2) at (-0.5,1);
    \fill (h2) circle (2pt) node[above=1pt]{$h_2$};
    \coordinate (h3) at (0.5,1);
    \fill (h3) circle (2pt) node[above=1pt]{$h_3$};
    \coordinate (h4) at (1.5,1);
    \fill (h4) circle (2pt) node[above=1pt]{$h_4$};
    \draw [thick] (g) -- (h4) (g) -- (h2) -- (h3) .. controls (-0.5,1.8) and (-0.5,1.8) .. (h1);
    \end{tikzpicture}}&=\left(\epsilon_{h_4}\cdot Y_{h_4}\right)\left(\epsilon_{h_3}\cdot F_{h_2}\cdot Y_{h_2}\right)\left(\epsilon_{h_1}\cdot k_{h_3}\right)
\end{align}
\endgroup

In the above discussion, we have given the graphic rule to read out the corresponding coefficients of a given KK basis. In Eq.~\eqref{eq:hmexpansion}, there are  $m!$ coefficients $C_{\pmb{\rho}}(\pmb{\sigma})$, each of which contains $m!$ terms given by $m!$ increasing trees. Thus naively we would expect that there are in all $(m!)^{2}$ trees involved. However, there are only $(m+1)^{m-1}$ distinct spanning trees for $m+1$ vertices ($m$ graviton vertices and a gluon root). Since $(m!)^{2}>(m+1)^{m-1}$ for $m\geqslant 2$, some trees must contribute to multiple $C_{\pmb{\rho}}(\pmb{\sigma})$ coefficients. The next problem we want to address is that given a spanning tree, how to find out all the $C_{\pmb{\rho}}(\pmb{\sigma})$ coefficients it can contribute to.
\begin{figure}[t]
  \centering
  \begin{tikzpicture}[vertex/.style={circle,draw,fill,inner sep=2pt},decoration={markings,mark=at position 0.5 with {\arrow{latex}}},edge/.style={very thick,postaction={decorate}}]
    \draw [dashed,blue] (-4.5,0.375) -- (4.5,0.375) -- (4.5,6.5) -- (-4.5,6.5) -- cycle;
    \node at (-4.5,6.5) [below right=2pt,text=blue]{level 1};
    \node (root) at (0,0) [vertex]{};
    \node at (root.south) [below=0pt]{root};
    \node (a1) at (150:1.5) [vertex]{};
    \node at (a1.west) [left=0pt]{$a_1$};
    \draw [edge] (root) -- (a1);
    \node (b1) at (30:1.5) [vertex]{};
    \node at (b1.east) [right=0pt]{$b_1$};
    \draw [edge] (root) --(b1);
    \draw [edge] (a1) -- ++(0,0.75) node (a2) [vertex]{};
    \draw [edge] (a2) -- ++(0,0.75) node (a3) [vertex]{};
    \node at (a2.west) [left=0pt]{$a_2$};
    \node at (a3.west) [left=0pt]{$a_3$};
    \draw [edge] (b1) -- ++(0,0.75) node (b2) [vertex]{};
    \draw [edge] (b2) -- ++(0,0.75) node (b3) [vertex]{};
    \node at (b2.east) [right=0pt]{$b_2$};
    \node at (b3.east) [right=0pt]{$b_3$};
    \node (c1) at (a1 -| root) [vertex]{};
    \node at (c1.west) [left=0pt]{$c_1$};
    \draw [edge] (root) -- (c1);
    \draw [edge] (c1) -- ++(0,0.75) node (c2) [vertex]{};
    \draw [edge] (c2) -- ++(0,0.75) node (c3) [vertex]{};
    \node at (c2.west) [left=0pt]{$c_2$};
    \node at (c3.west) [left=0pt]{$c_3$};
    \draw [dashed,red] (c3) ++(135:0.5) -- ++(4.5,0) -- ++(0,3.5) -- ++(-7,0) node [below right=2pt,text=red]{level 2} -- ++(0,-3.5) -- cycle;
    \draw [edge] (c3) -- ++(135:1) node (d1) [vertex]{};
    \draw [edge] (d1) -- ++(135:1) node (d2) [vertex]{};
    \node at (d1.north east) [above=0pt]{$d_1$};
    \node at (d2.north east) [above=0pt]{$d_2$};
    \draw [edge] (c3) -- ++(45:1) node (e1) [vertex]{};
    \draw [edge] (e1) -- ++(45:1) node (e2) [vertex]{};
    \node at (e1.north) [above=0pt]{$e_1$};
    \node at (e2.south east) [below=0pt]{$e_2$};
    \draw [purple,dashed] (e2) ++(135:0.5) -- ++(2.5,0) -- ++(0,1.8) -- ++(-4.5,0) node [below right=1pt]{level 3} -- ++(0,-1.8) -- cycle;
    \draw [edge] (e2) -- ++(135:1) node (f1) [vertex]{};
    \node at (f1.west) [left=0pt]{$f_1$};
    \draw [edge] (f1) -- ++(135:1) node (f2) [vertex]{};
    \node at (f2.west) [left=0pt]{$f_2$};
    \draw [edge] (e2) -- ++(45:1) node (g1) [vertex]{};
    \draw [edge] (g1) -- ++(45:1) node (g2) [vertex]{};
    \node at (g1.east) [right=0pt]{$g_1$};
    \node at (g2.east) [right=0pt]{$g_2$};
  \end{tikzpicture}
  \caption{The rooted tree with three levels. }
  \label{Tree}
\end{figure}
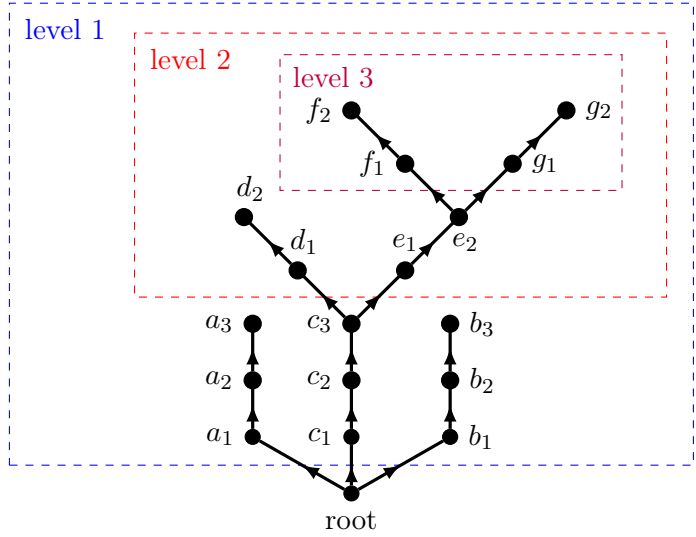

The answer to this question is simple: \emph{suppose we define the set of all the increasing trees respecting the order $\pmb{\sigma}$ as $IT(\pmb{\sigma})$, then given a spanning tree $\mathsf{T}$, it will contribute to all those $\pmb{\sigma}$'s whose $IT(\pmb{\sigma})$ contains $\mathsf{T}$.} These $\pmb{\sigma}$'s can be read out in the following way. First let us define the level structure of a spanning tree, demonstrated by Figure~\ref{Tree}:
\begin{itemize}

\item From the root, there are three outgoing arrows, so we have three subsets in the first level.

\item The subsets of $a$ and $b$ are simple, which are $\{a_1, a_2,a_3\}$ and $\{ b_1, b_2, b_3\}$ respectively.

\item For the middle subset, it is much more complicated. Along the arrow, first we have $\{c_1, c_2, c_3\}$. However, there are two outgoing branches starting from $c_3$. Thus we have a second level with two branches: one is $\{d_1, d_2\}$ and the other is $\{e_1, e_2\}$. Again, starting from $e_2$ there are two
branches: $\{f_1, f_2\}$ and $\{g_1, g_2\}$, which give us the
third level.

\item Putting everything together, we have the following level structure:
\bea \left\{\begin{array}{l} \{a_1, a_2, a_3\} \\
 \{ b_1, b_2, b_3\} \\  \left\{c_1, c_2, c_3,
\begin{array}{l} \{d_1, d_2\} \\
\left\{e_1, e_2,
\left\{\begin{array}{l} \{f_1,
f_2\} \\ \{g_1, g_2\} \end{array}\right.
\right\}
\end{array}
\right\} \end{array} \right\} \eea

\end{itemize}
After reading out the level structure, we can write down the orderings each level gives by shuffle actions. First, there are two branches at level $3$, such that this level gives:
\begin{equation*}
  {\tt level3}
  =\{f_1,f_2\}\shuffle\{g_1,g_2\}\,,
\end{equation*}
Then at the second level, there are also two branches. The first is simply $\{d_1,d_2\}$, while the second one is $\{e_1,e_2\}$ concatenating {\tt level3}. The orderings given by this level are:
\begin{equation*}
  {\tt level2}=\{d_1,d_2\}\shuffle\{e_1,e_2,{\tt level3}\}\,.
\end{equation*}
Finally, at the first level, we have:
\begin{equation*}
  {\tt level1}=\{a_1,a_2,a_3\}\shuffle\{b_1,b_2,b_3\}\shuffle\{c_1,c_2,c_3,{\tt level2}\}\,.
\end{equation*}
After the shuffle products are fully expanded, {\tt level1} gives the set of orderings this tree contributes to.

Next, we check this rule with our three-graviton expressions. In Eq.~\eqref{C123-tree}, the first tree only has one level, such that
\begin{align}
\adjustbox{raise=-0.75cm}{\begin{tikzpicture}
    \coordinate (g) at (0,0);
    \fill (g) circle (2pt) node[below=1pt]{$\mathsf{g}$};
    \coordinate (h1) at (-1,1);
    \fill (h1) circle (2pt) node[above=1pt]{$h_1$};
    \coordinate (h2) at (0,1);
    \fill (h2) circle (2pt) node[above=1pt]{$h_2$};
    \coordinate (h3) at (1,1);
    \fill (h3) circle (2pt) node[above=1pt]{$h_3$};
    \draw [thick] (g) -- (h1) (g) -- (h2) (g) -- (h3);
\end{tikzpicture}}=\left(\epsilon_{h_1}\cdot Y_{h_1}\right)\left(\epsilon_{h_1}\cdot Y_{h_1}\right)\left(\epsilon_{h_1}\cdot Y_{h_1}\right)& & \begin{array}{l}
  \text{contributes to} \\
  \quad \{h_1\}\shuffle\{h_2\}\shuffle\{h_3\} \\
  =\{h_1h_2h_3\}+\{h_1h_3h_2\}+\{h_2h_1h_3\} \\
  \quad+\;\{h_2h_3h_1\}+\{h_3h_1h_2\}+\{h_3h_2h_1\} \\
\end{array}
\end{align}
Indeed, this term is contained in all six $C_{\pmb{\rho}}(\pmb{\sigma})$, given in Eq.~\eqref{eq:C123} to \eqref{eq:C312}. For tree 5 in Eq.~\eqref{C123-tree}, there is only one branch at level one but two branches at level two, such that
\begin{align}
\adjustbox{raise=-0.75cm}{\begin{tikzpicture}
    \coordinate (g) at (0,0);
    \fill (g) circle (2pt) node[below=1pt]{$\mathsf{g}$};
    \coordinate (h1) at (-1,1);
    \fill (h1) circle (2pt) node[above=1pt]{$h_1$};
    \coordinate (h2) at (0,1);
    \fill (h2) circle (2pt) node[above=1pt]{$h_2$};
    \coordinate (h3) at (1,1);
    \fill (h3) circle (2pt) node[above=1pt]{$h_3$};
    \draw [thick] (g) -- (h1) -- (h2) (g) (h1) .. controls (0,1.8) and (0,1.8) .. (h3);
\end{tikzpicture}}=\left(\epsilon_{h_3}\cdot F_{h_1}\cdot Y_{h_1}\right)\left(\epsilon_{h_2}\cdot k_{h_1}\right)& & \begin{array}{l}
  \text{contributes to} \\
  \{h_1,\{h_2\}\shuffle\{h_3\}\}=\{h_1h_2h_3\}+\{h_1h_3h_2\}
\end{array}
\end{align}
This expression can only be found in $C_{[321]}(123)$ and $C_{[321]}(132)$ as expected. Last but not least, the tree in Eq.~\eqref{321-tree} only has one level, contributing only to the order $\{321\}$. Again, the corresponding term can only be found in $C_{[321]}(321)$.

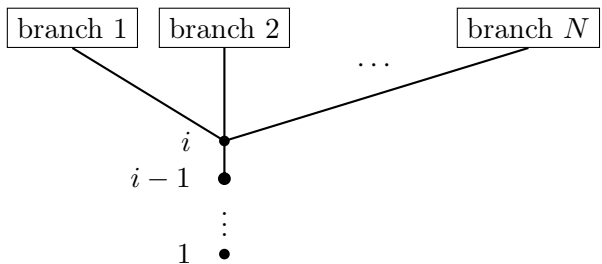
\begin{figure}[t]
  \centering
    \begin{tikzpicture}
    \coordinate (o) at (0,0);
    \fill (o) circle (2pt) node[left=3mm]{$i$};
    \node (t1) at (-2,1.5) [draw,rectangle]{branch $1$};
    \node (t2) at (0,1.5) [draw,rectangle]{branch $2$};
    \node at (2,1) [] {$\cdots$};
    \node (t3) at (4,1.5) [draw,rectangle]{branch $N$};
    \draw [thick] (o) --  (t1.south) (o) -- (t2) (o) -- (t3.south);
    \draw [thick,fill] (o) -- ++(0,-0.5) circle (2pt) node[left=3mm]{$i-1$};
    \node at (0,-1) {$\vdots$};
    \fill (0,-1.5) circle (2pt) node[left=3mm]{$1$};
    \end{tikzpicture}
    \caption{A tree whose first level contains $N$ branches.}
    \label{level-tree}
\end{figure}

%
%
%
%
Using the above understanding, we also give two four-graviton examples:
\begingroup
\allowdisplaybreaks
\begin{align}
    &\adjustbox{raise=-0.75cm}{\begin{tikzpicture}
    \coordinate (g) at (0,0);
    \fill (g) circle (2pt) node[below=1pt]{$\mathsf{g}$};
    \coordinate (h1) at (-1.5,1);
    \fill (h1) circle (2pt) node[above=1pt]{$h_1$};
    \coordinate (h2) at (-0.5,1);
    \fill (h2) circle (2pt) node[above=1pt]{$h_2$};
    \coordinate (h3) at (0.5,1);
    \fill (h3) circle (2pt) node[above=1pt]{$h_3$};
    \coordinate (h4) at (1.5,1);
    \fill (h4) circle (2pt) node[above=1pt]{$h_4$};
    \draw [thick] (g) -- (h4) -- (h3) -- (h2) (h1) .. controls (0,2) and (0,2) .. (h4);
    \end{tikzpicture}}& &\text{contributes to}& & \begin{array}{l}
        \quad\{h_4, \{h_1\}\shuffle\{h_3h_2\}~\}\\
        =\{h_4h_1h_3h_2\}+\{h_4h_3h_1h_2\}+\{h_4h_3h_2h_1\}
    \end{array} \nonumber\\
    &\adjustbox{raise=-0.75cm}{\begin{tikzpicture}
    \coordinate (g) at (0,0);
    \fill (g) circle (2pt) node[below=1pt]{$\mathsf{g}$};
    \coordinate (h1) at (-1.5,1);
    \fill (h1) circle (2pt) node[above=1pt]{$h_1$};
    \coordinate (h2) at (-0.5,1);
    \fill (h2) circle (2pt) node[above=1pt]{$h_2$};
    \coordinate (h3) at (0.5,1);
    \fill (h3) circle (2pt) node[above=1pt]{$h_3$};
    \coordinate (h4) at (1.5,1);
    \fill (h4) circle (2pt) node[above=1pt]{$h_4$};
    \draw [thick] (g) -- (h3) -- (h4) (g) -- (h2) -- (h1);
    \end{tikzpicture}}& &\text{contributes to}& & \begin{array}{l}
        \quad\{h_2h_1\}\shuffle\{h_3h_4\} \\
        =\{h_2h_1h_3h_4\}+\{h_2h_3h_1h_4\}+\{h_2h_3h_4h_1\} \\
        \quad+\;\{h_3h_4h_2h_1\}+\{h_3h_2h_4h_1\}+\{h_3h_2h_1h_4\}
    \end{array}
\end{align}
\endgroup
Before closing this section, we note that the above rules can be recursively defined for generic trees. First, the level splitting happens only when {a branch in the tree splits}. Then for a subtree with the structure of Figure~\ref{level-tree}, it will contribute to the following set of orderings:
\begin{equation}
  \{1,2\ldots i,{\tt Branch1}\shuffle{\tt Branch2}\shuffle\ldots\shuffle{\tt BranchN}\}\,,
\end{equation}
where, for example, {\tt Branch1} is the set of orderings contributed by the first branch, etc, which can be recursively constructed using the above relation, until the highest level.

\section{Conclusion and discussion}\label{sec:conclusion}

In this paper, we have established a recursive expansion \eqref{eq:hmresult} for tree level single trace EYM amplitudes with $m$ gravitons and $n$ gluons, facilitated by expanding the CHY integrand $\Pf(\Psi_{\mathsf{H}})$ along one row. It works for arbitrary number of gravitons and gluons, as we have proved in Section~\ref{sec:recursive}. This recursive construction agrees with the one proposed in a recent paper~\cite{Fu:2017uzt}, which is observed simply by imposing gauge invariance. Our paper gives a firm proof of the observation. We note that our recursive expansion holds only on-shell, namely, the momentum conservation, transversal condition and scattering equations should all be satisfied.

Very remarkably, once we fully expand the $m$-graviton and $n$-gluon EYM amplitudes in terms of pure YM amplitudes with $n+m$ gluons in the KK basis, all the expansion coefficients can be written down through a set of elegant graph theory rules. To be specific, the coefficient of each KK basis amplitude contains exactly $m!$ terms, each of which corresponds to an increasing tree with respect to the order of the gravitons in a certain KK basis amplitude. Then the evaluation of each tree relies on a reference order. On one hand, the origin of this reference order is how we expand $\Pf(\Psi_{\mathsf{H}})$ step by step. On the other hand, it serves as a ``gauge choice'' in the over-complete KK basis YM amplitudes.

From these results, one very promising direction is of course the construction BCJ numerators~\cite{Bern:2008qj,Bern:2010ue}. It is pointed out in~\cite{Fu:2017uzt} that the YM integrand $\Pfp(\Psi)$ can be readily expanded in terms of the EYM integrands with polynomial coefficients. Then our results essentially provides a working algorithm to write down the polynomial BCJ numerators for tree level YM, especially the spanning tree structure which has not been
explicitly spelled out in this paper. This study may shed lights on new understandings of the numerator algebra, and also on possible new formulations of YM that manifest the numerator algebra. Such a formulation already exists for NLSM~\cite{Cheung:2016prv}, and it will be fascinating if it also exists for YM.

Our work shows that a controlled expansion of CHY integrands can produce very fruitful results. It will be interesting to see, for example, how the expansion of $\Pfp(A)$, the CHY integrand for NLSM\footnote{{Such an expansion of the NLSM integrand should give us a set of BCJ numerators in the KK basis. It is also interesting to compare this result with known forms of the NLSM BCJ numerators~\cite{Du:2016tbc,Carrasco:2016ldy,Carrasco:2016ygv}.}}, interplays with~\cite{Cheung:2016prv}. One can also try the CHY integrand at the one-loop level for YM theory \cite{Geyer:2015bja,Geyer:2015jch,Cachazo:2015aol}. Another topic that worths more attention is the study of multitrace EYM integrands~\cite{Cachazo:2014xea,Nandan:2016pya,Schlotterer:2016cxa}, and its relation with the known results constructed from double copy~\cite{Chiodaroli:2014xia,Chiodaroli:2015rdg}.

{\bf Note added} During the final completion of this work, we came aware of~\cite{Chiodaroli:2017ngp}, which partially overlaps with our results.

\acknowledgments

We would like to thank Yi-Jian Du, Chih-Hao Fu and Rijun Huang for
valuable discussions. BF is supported by Qiu-Shi Funding and the
National Natural Science Foundation of China (NSFC) with Grant
No.11575156, No.11135006 and No.11125523.

\appendix

\section{Understanding the graphic rules}\label{sec:derivation}
\begin{figure}[t]
    \centering
    \begin{tikzpicture}[vertex/.style={draw,thick,rectangle,rounded corners}]
    \node (a) at (0,0) [vertex]{$\Pf(\Psi_{m})\PT(\mathsf{G})$};
    \node (b) at (3.5,2) [vertex,align=center]{$\left(\epsilon_{h_m}\cdot Y_{h_m}\right)\Pf(\Psi_{m-1})$ \\ $\times\PT(\mathsf{G},h_m)$};
    \node (c) at (3.5,-1) [vertex]{$\epsilon_{h_m}\cdot F_{h_s}\cdot T[h_sh_m]$};
    \node (d) at (8,0) [vertex,align=center]{$\left(\epsilon_{h_m}\cdot F_{h_s}\cdot Y_{h_s}\right)\Pf(\Psi_{m-2})$ \\ $\times\PT(\mathsf{G},h_s,h_m)$};
    \node (e) at (8,-2) [vertex]{$\epsilon_{h_m}\cdot F_{h_s}\cdot F_{h_t}\cdot T[h_th_sh_m]$};
    \node at (e.east) [right=0.5cm] {$\cdots$};
    \node at (d.east) [right=0.5cm] {$\cdots$};
    \node at (b.east) [right=0.5cm] {$\cdots$};
    \draw [-stealth,rounded corners,very thick,red] (a.south) -- (0,-1) -- (c.west);
    \draw [-stealth,rounded corners,very thick,red] (c.south) -- (3.5,-2) -- (e.west);
    \draw [-stealth,rounded corners,very thick,blue] (a.north) -- (0,2) -- (b.west);
    \draw [-stealth,rounded corners,very thick,blue] (c.north) -- (3.5,0) -- (d.west);
    \end{tikzpicture}
    \caption{Schematic expansion of the EYM integrand}
    \label{fig:expansion}
\end{figure}
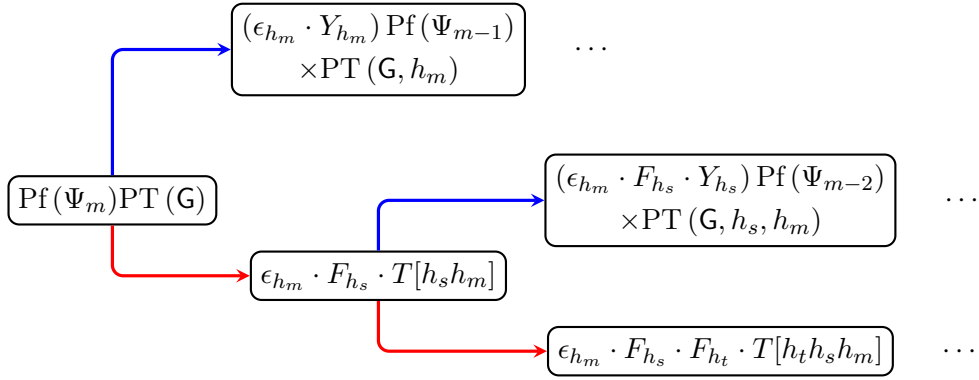

This appendix is devoted to show how our graphic rules given in Section~\ref{sec:GraphicRules} emerge from our recursive relation \eqref{eq:hmresult}. Suppose we are about to expand the $m$-graviton
integrand
\begin{equation*}
(-1)^{m}\Pf(\Psi_{\mathsf{H}})\,\PT(12\ldots n)\,,
\end{equation*}
we have to make a choice on along which row do we expand the Pfaffian. Throughout our discussion in this paper, we have consistently chosen $h_m$, namely,
\begin{equation}
\label{choice1}
(-1)^{m}\Pf(\Psi_{\mathsf{H}})\,\PT(12\ldots n)=\epsilon_{h_m}\cdot T\left[1,\{2\ldots n-1\}\shuffle\{h_m\},n\,|\,\{h_{m-1}\ldots h_1\}\right]\,.
\end{equation}
Of course the value of this integrand will not change should we
choose any other graviton to perform the expansion, although the
final pure YM expansion coefficients may have a different form. Now
we carry out the expansion one step further, according to
Eq.~\eqref{eq:hmresult}:
\begin{align}
&\quad\;\epsilon_{h_m}\cdot T\left[1,\{2\ldots n-1\}\shuffle\{h_m\},n\,|\,\{h_{m-1}\ldots h_1\}\right]\nonumber\\
&=\sum_{\shuffle}(\epsilon_{h_m}\cdot Y_{h_m})\,\Pf(\Psi_{h_{m-1}\ldots h_{1}}^{n+m})\,\PT(1,\{2\ldots n-1\}\shuffle\{h_m\},n)\nonumber\\
&\quad+(-1)^{m-1}\sum_{s=1}^{m-1}\epsilon_{h_m}\cdot F_{h_s}
\cdot T[1,\{2\ldots n-1\}\shuffle\{h_sh_m\},n\,|\,\{h_{m-1}\ldots\slashed{h}_s\ldots h_1\}]\,.
\end{align}
In the first term, the factor $\left(\epsilon_{h_m}\cdot Y_{h_m}\right)$ can  be viewed as connecting the graviton vertex
$h_m$ to the root, while in the second term, $\left(\epsilon_{h_m}\cdot F_{h_s}\right)^{\mu}$ can be interpreted as connecting $h_m$ to another graviton $h_s$, since in the ordering $\{h_sh_m\}$, $h_s$ is at the left
hand side of the $h_m$. For the second term, further expansion
will again lead to two term, one with $F_{h_s}$ connected to the
root, giving $\left(\epsilon_{h_m}\cdot F_{h_s}\cdot Y_{h_s}\right)$
and an $(m-2)$-graviton EYM integrand:
$$\sum_{\shuffle}\left(\epsilon_{h_m}\cdot F_{h_s}\cdot Y_{h_s}\right)\Pf(\Psi^{n+m}_{h_{m-1}\ldots\slashed{h}_s\ldots h_1})\,\PT(1,\{2\ldots n-1\}\shuffle\{h_sh_m\},n)\,, $$
and the other with $F_{h_s}$ connected to yet another graviton, giving
$$\epsilon_{h_m}\cdot F_{h_s}\cdot F_{h_t}\cdot T\left[1,\{2\ldots n-1\}\shuffle\{h_th_sh_m\},n\,|\,\{h_{m-1}\ldots\slashed{h}_s\ldots\slashed{h}_t\ldots h_1\}\right]\,.$$
We can do it recursively and this procedure gives,
in fact, a construction of those \emph{increasing trees} presented in {\bf
Step 1} in the Section~\ref{sec:GraphicRules}. Furthermore, the starting vertex $h_m$ plays the role of $\rho(1)$ in the reference order ${\pmb \rho}$. For convenience, we show this expansion process
schematically in Figure~\ref{fig:expansion}.

Now we work out more details for the above construction by an example. We first keep following the red arrows in Figure~\ref{fig:expansion} to connect more gravitons until the {vertex} $\phi_\ell$, which is then connected to the gluon root. In this way, we get a chain evaluated as
\begin{equation}
\adjustbox{raise=-0.5cm}{\begin{tikzpicture}
    \draw [thick,fill] (0,0) circle (2pt) node[below=1pt]{$\mathsf{g}$} -- (1,0) circle (2pt) node[below=1pt]{$\phi_{\ell}$} -- (2,0) circle (2pt) node[below=1pt]{$\phi_{\ell-1}$} (2.5,0) node {$\cdots$} (3,0) circle (2pt) node[below=1pt]{$\phi_2$} -- (4,0) circle (2pt) node[below=1pt]{$\phi_1$};
    \end{tikzpicture}}=\left(\epsilon_{\phi_1}\cdot F_{\phi_2}\cdots F_{\phi_m}\cdot Y_{\phi_m}\right)\,,
\end{equation}
where we have defined the first {vertex} $\phi_1\equiv h_m$. In the scheme of Figure~\ref{fig:expansion}, terminating a graviton chain at the gluon root essentially means that we switch to a blue arrow and arrive at the block
$$ \sum_{\shuffle}\left(\epsilon_{\phi_1}\cdot F_{\phi_2}\cdots F_{\phi_m}\cdot Y_{\phi_m}\right)\Pf(\Psi^{n+m}_{\theta_1\ldots\theta_t})\,\PT(1,\{2\ldots n-1\}\shuffle\{\phi_\ell\ldots\phi_1\},n)\,,$$
where $\{\theta_1\ldots\theta_t\}$ is the complement of $\{\phi_1\ldots\phi_\ell\}$ in $\mathsf{H}=\{h_1\ldots h_m\}$. Now we come across a Pfaffian again, and we have to choose another graviton along which we expand this Pfaffian. Thus here comes the origin of the reference order $\pmb{\rho}$: \emph{it is a list of priority we give to each graviton, such that once we need to expand a Pfaffian, we always choose to expand along the one with the highest priority according to $\pmb\rho$.} In this case, we simply delete $\{\phi_1\ldots\phi_\ell\}$ from $\pmb\rho$ and pick up the first remaining graviton, say $\theta_1$. We then stay with the red arrows until we deplete all the gravitons, which gives us another chain. The final block we reach gives:
\begin{align}
&\sum_{\shuffle_1}\sum_{\shuffle_2}\left(\epsilon_{\phi_1}\cdot F_{\phi_2}\cdots F_{\phi_\ell}\cdot Y_{\phi_\ell}\right)\left(\epsilon_{\theta_1}\cdot F_{\theta_2}\cdots F_{\theta_t}\cdot X_{\theta_t}\right)\PT(1,\{2\ldots n-1\}\shuffle_1\{\phi_{\ell}\ldots \phi_1\}\shuffle_2\{\theta_t\ldots \theta_1\})\,.
\end{align}
Since we have treated $\{\phi_\ell\ldots\phi_1\}$ as gluons when expanding the $\theta$-chain, we need to include $k_{\phi_i}$ into $X_{\theta_t}$ if there are some $\phi_i$ before $\theta_t$:
\begin{equation}
X_{\theta_t}=Y_{\theta_t}+\sum_{\phi_i\prec\theta_t}k_{\phi_i}\,.
\end{equation}
After fully expanding the shuffle $\{\phi_\ell\ldots\phi_1\}\shuffle\{\theta_t\ldots\theta_1\}$, one can notice that the factor $\left(\epsilon_{\theta_1}\cdot F_{\theta_2}\cdots F_{\theta_t}\cdot Y_{\theta_t}\right)$ is contained in the coefficients of all the corresponding KK basis:
\begin{align}
\label{eq:tree1}
\left(\epsilon_{\phi_1}\cdot F_{\phi_2}\cdots F_{\phi_\ell}\cdot Y_{\phi_\ell}\right)\left(\epsilon_{\theta_1}\cdot F_{\theta_2}\cdots F_{\theta_t}\cdot Y_{\theta_t}\right) \subset \{\phi_\ell\ldots\phi_1\}\shuffle\{\theta_t\ldots\theta_1\}\,.
\end{align}
In addition, the factor $\left(\epsilon_{\theta_1}\cdot F_{\theta_2}\cdots F_{\theta_t}\cdot k_{\phi_\ell}\right)$ appears when $\phi_\ell$ is ahead of $\theta_t$:
\begin{align}
\left(\epsilon_{\phi_1}\cdot F_{\phi_2}\cdots F_{\phi_\ell}\cdot Y_{\phi_\ell}\right)\left(\epsilon_{\theta_1}\cdot F_{\theta_2}\cdots F_{\theta_t}\cdot k_{\phi_\ell}\right) \subset \left\{\phi_\ell,\{\phi_{\ell-1}\ldots \phi_1\}\shuffle\{\theta_t\ldots\theta_1\}\right\}\,.
\end{align}
Generally, we have:
\begin{align}
\label{eq:tree2}
\left(\epsilon_{\phi_1}\cdot F_{\phi_2}\cdots F_{\phi_\ell}\cdot Y_{\phi_\ell}\right)\left(\epsilon_{\theta_1}\cdot F_{\theta_2}\cdots F_{\theta_t}\cdot k_{\phi_{i}}\right) \subset \left\{\phi_\ell\ldots\phi_{i},\{\phi_{i-1}\ldots \phi_1\}\shuffle\{\theta_t\ldots\theta_1\}\right\}\,.
\end{align}
Now one can understand how the structure of increasing tree emerges. We can graphically represent Eq.~\eqref{eq:tree1} to \eqref{eq:tree2} as follows:
\begin{align}
\label{eq:treerule1}
\adjustbox{raise=-1cm}{\begin{tikzpicture}
    \draw [thick,fill] (0,0) circle (2pt) node[below=1pt]{$\mathsf{g}$} -- (1,0) circle (2pt) node[below=1pt]{$\phi_{\ell}$} -- (2,0) circle (2pt) node[below=1pt]{$\phi_{\ell-1}$} (2.5,0) node {$\cdots$} (3,0) circle (2pt) node[below=1pt]{$\phi_2$} -- (4,0) circle (2pt) node[below=1pt]{$\phi_1$};
    \draw [thick,fill] (0,0) -- (1,1) circle (2pt) node[above=1pt]{$\theta_t$} -- (2,1) circle (2pt) node[above=1pt]{$\theta_{t-1}$} (2.5,1) node {$\cdots$} (3,1) circle (2pt) node[above=1pt]{$\theta_2$} -- (4,1) circle (2pt) node[above=1pt]{$\theta_1$};
    \end{tikzpicture}}&=\left(\epsilon_{\phi_1}\cdot F_{\phi_2}\cdots F_{\phi_m}\cdot Y_{\phi_\ell}\right)\left(\epsilon_{\theta_1}\cdot F_{\theta_2}\cdots F_{\theta_t}\cdot Y_{\theta_t}\right)\,,\\
\label{eq:treerule2}
\adjustbox{raise=-1cm}{\begin{tikzpicture}
    \draw [thick,fill] (0,0) circle (2pt) node[below=1pt]{$\mathsf{g}$} -- (1,0) circle (2pt) node[below=1pt]{$\phi_{\ell}$} (1.5,0) node{$\cdots$} (2,0) circle (2pt) node[below=1pt]{$\phi_{i}$} (2.5,0) node {$\cdots$} (3,0) circle (2pt) node[below=1pt]{$\phi_2$} -- (4,0) circle (2pt) node[below=1pt]{$\phi_1$};
    \draw [thick,fill] (2,0) -- (1,1) circle (2pt) node[above=1pt]{$\theta_t$} -- (2,1) circle (2pt) node[above=1pt]{$\theta_{t-1}$} (2.5,1) node {$\cdots$} (3,1) circle (2pt) node[above=1pt]{$\theta_2$} -- (4,1) circle (2pt) node[above=1pt]{$\theta_1$};
    \end{tikzpicture}}&=\left(\epsilon_{\phi_1}\cdot F_{\phi_2}\cdots F_{\phi_m}\cdot Y_{\phi_\ell}\right)\left(\epsilon_{\theta_1}\cdot F_{\theta_2}\cdots F_{\theta_t}\cdot k_{\phi_i}\right)\,.
\end{align}
It is interesting to observe that the tree in \eqref{eq:treerule1} is an increasing tree respecting any $\pmb{\sigma}\in\{\phi_\ell\ldots \phi_1\}\shuffle\{\theta_t\ldots \theta_1\}$. Similarly, the tree in \eqref{eq:treerule2} is an increasing tree respecting any $\pmb{\sigma}\in\{\phi_\ell\ldots\phi_i,\{\phi_{i-1}\ldots\phi_1\}\shuffle\{\theta_t\ldots\theta_1\}\}$. Our derivation also shows that the terms evaluated from Eq.~\eqref{eq:treerule1} and \eqref{eq:treerule2} do contribute to all such $\pmb{\sigma}$'s represented by the two trees.
Finally, this example shows how we should close a chain that ends on a graviton vertex, as given in {\bf Step 2} in Section~\ref{sec:GraphicRules}

The above two-chain calculation demonstrates how the graph theory rules presented in Section~\ref{sec:GraphicRules}, in their primitive forms, originates from our recursive relation. Of course, we have already refined these rules into a set of algorithms  in Section~\ref{sec:GraphicRules} such that they can be applied to the most generic cases.

\bibliographystyle{JHEP}
\bibliography{Refs}

\providecommand{\href}[2]{#2}\begingroup\raggedright\begin{thebibliography}{10}

\bibitem{Stieberger:2016lng}
S.~Stieberger and T.~R. Taylor, \emph{{New relations for Einstein-Yang-Mills
  amplitudes}},
  \href{http://dx.doi.org/10.1016/j.nuclphysb.2016.09.014}{\emph{Nucl. Phys.}
  {\bfseries B913} (2016) 151--162},
  [\href{https://arxiv.org/abs/1606.09616}{{\ttfamily 1606.09616}}].

\bibitem{Nandan:2016pya}
D.~Nandan, J.~Plefka, O.~Schlotterer and C.~Wen, \emph{{Einstein-Yang-Mills
  from pure Yang-Mills amplitudes}},
  \href{http://dx.doi.org/10.1007/JHEP10(2016)070}{\emph{JHEP} {\bfseries 10}
  (2016) 070}, [\href{https://arxiv.org/abs/1607.05701}{{\ttfamily
  1607.05701}}].

\bibitem{delaCruz:2016gnm}
L.~de~la Cruz, A.~Kniss and S.~Weinzierl, \emph{{Relations for
  Einstein-Yang-Mills amplitudes from the CHY representation}},
  \href{http://dx.doi.org/10.1016/j.physletb.2017.01.036}{\emph{Phys. Lett.}
  {\bfseries B767} (2017) 86--90},
  [\href{https://arxiv.org/abs/1607.06036}{{\ttfamily 1607.06036}}].

\bibitem{Schlotterer:2016cxa}
O.~Schlotterer, \emph{{Amplitude relations in heterotic string theory and
  Einstein-Yang-Mills}},
  \href{http://dx.doi.org/10.1007/JHEP11(2016)074}{\emph{JHEP} {\bfseries 11}
  (2016) 074}, [\href{https://arxiv.org/abs/1608.00130}{{\ttfamily
  1608.00130}}].

\bibitem{Du:2016wkt}
Y.-J. Du, F.~Teng and Y.-S. Wu, \emph{{Direct Evaluation of $n$-point
  single-trace MHV amplitudes in 4d Einstein-Yang-Mills theory using the CHY
  Formalism}}, \href{http://dx.doi.org/10.1007/JHEP09(2016)171}{\emph{JHEP}
  {\bfseries 09} (2016) 171},
  [\href{https://arxiv.org/abs/1608.00883}{{\ttfamily 1608.00883}}].

\bibitem{Nandan:2016ohb}
D.~Nandan, J.~Plefka and W.~Wormsbecher, \emph{{Collinear limits beyond the
  leading order from the scattering equations}},
  \href{http://dx.doi.org/10.1007/JHEP02(2017)038}{\emph{JHEP} {\bfseries 02}
  (2017) 038}, [\href{https://arxiv.org/abs/1608.04730}{{\ttfamily
  1608.04730}}].

\bibitem{He:2016mzd}
S.~He and O.~Schlotterer, \emph{{Loop-level KLT, BCJ and EYM amplitude
  relations}},  \href{https://arxiv.org/abs/1612.00417}{{\ttfamily
  1612.00417}}.

\bibitem{Fu:2017uzt}
C.-H. Fu, Y.-J. Du, R.~Huang and B.~Feng, \emph{{Expansion of
  Einstein-Yang-Mills Amplitude}},
  \href{https://arxiv.org/abs/1702.08158}{{\ttfamily 1702.08158}}.

\bibitem{Cachazo:2013iaa}
F.~Cachazo, S.~He and E.~Y. Yuan, \emph{{Scattering in Three Dimensions from
  Rational Maps}}, \href{http://dx.doi.org/10.1007/JHEP10(2013)141}{\emph{JHEP}
  {\bfseries 10} (2013) 141},
  [\href{https://arxiv.org/abs/1306.2962}{{\ttfamily 1306.2962}}].

\bibitem{Cachazo:2013gna}
F.~Cachazo, S.~He and E.~Y. Yuan, \emph{{Scattering equations and
  Kawai-Lewellen-Tye orthogonality}},
  \href{http://dx.doi.org/10.1103/PhysRevD.90.065001}{\emph{Phys. Rev.}
  {\bfseries D90} (2014) 065001},
  [\href{https://arxiv.org/abs/1306.6575}{{\ttfamily 1306.6575}}].

\bibitem{Cachazo:2013hca}
F.~Cachazo, S.~He and E.~Y. Yuan, \emph{{Scattering of Massless Particles in
  Arbitrary Dimensions}},
  \href{http://dx.doi.org/10.1103/PhysRevLett.113.171601}{\emph{Phys.Rev.Lett.}
  {\bfseries 113} (2014) 171601},
  [\href{https://arxiv.org/abs/1307.2199}{{\ttfamily 1307.2199}}].

\bibitem{Cachazo:2013iea}
F.~Cachazo, S.~He and E.~Y. Yuan, \emph{{Scattering of Massless Particles:
  Scalars, Gluons and Gravitons}},
  \href{http://dx.doi.org/10.1007/JHEP07(2014)033}{\emph{JHEP} {\bfseries 1407}
  (2014) 033}, [\href{https://arxiv.org/abs/1309.0885}{{\ttfamily 1309.0885}}].

\bibitem{Cachazo:2014nsa}
F.~Cachazo, S.~He and E.~Y. Yuan, \emph{{Einstein-Yang-Mills Scattering
  Amplitudes From Scattering Equations}},
  \href{http://dx.doi.org/10.1007/JHEP01(2015)121}{\emph{JHEP} {\bfseries 01}
  (2015) 121}, [\href{https://arxiv.org/abs/1409.8256}{{\ttfamily 1409.8256}}].

\bibitem{Cachazo:2014xea}
F.~Cachazo, S.~He and E.~Y. Yuan, \emph{{Scattering Equations and Matrices:
  From Einstein To Yang-Mills, DBI and NLSM}},
  \href{http://dx.doi.org/10.1007/JHEP07(2015)149}{\emph{JHEP} {\bfseries 07}
  (2015) 149}, [\href{https://arxiv.org/abs/1412.3479}{{\ttfamily 1412.3479}}].

\bibitem{Bjerrum-Bohr:2016juj}
N.~E.~J. Bjerrum-Bohr, J.~L. Bourjaily, P.~H. Damgaard and B.~Feng,
  \emph{{Analytic representations of Yang-Mills amplitudes}},
  \href{http://dx.doi.org/10.1016/j.nuclphysb.2016.10.012}{\emph{Nucl. Phys.}
  {\bfseries B913} (2016) 964--986},
  [\href{https://arxiv.org/abs/1605.06501}{{\ttfamily 1605.06501}}].

\bibitem{Cardona:2016gon}
C.~Cardona, B.~Feng, H.~Gomez and R.~Huang, \emph{{Cross-ratio Identities and
  Higher-order Poles of CHY-integrand}},
  \href{http://dx.doi.org/10.1007/JHEP09(2016)133}{\emph{JHEP} {\bfseries 09}
  (2016) 133}, [\href{https://arxiv.org/abs/1606.00670}{{\ttfamily
  1606.00670}}].

\bibitem{Huang:2017ydz}
R.~Huang, Y.-J. Du and B.~Feng, \emph{{Understanding the Cancelation of Double
  Poles in the Pfaffian of CHY-formulism}},
  \href{https://arxiv.org/abs/1702.05840}{{\ttfamily 1702.05840}}.

\bibitem{Boels:2016xhc}
R.~H. Boels and R.~Medina, \emph{{Graviton and gluon scattering from first
  principles}},
  \href{http://dx.doi.org/10.1103/PhysRevLett.118.061602}{\emph{Phys. Rev.
  Lett.} {\bfseries 118} (2017) 061602},
  [\href{https://arxiv.org/abs/1607.08246}{{\ttfamily 1607.08246}}].

\bibitem{Arkani-Hamed:2016rak}
N.~Arkani-Hamed, L.~Rodina and J.~Trnka, \emph{{Locality and Unitarity from
  Singularities and Gauge Invariance}},
  \href{https://arxiv.org/abs/1612.02797}{{\ttfamily 1612.02797}}.

\bibitem{Rodina:2016mbk}
L.~Rodina, \emph{{Uniqueness from locality and BCFW shifts}},
  \href{https://arxiv.org/abs/1612.03885}{{\ttfamily 1612.03885}}.

\bibitem{Rodina:2016jyz}
L.~Rodina, \emph{{Uniqueness from gauge invariance and the Adler zero}},
  \href{https://arxiv.org/abs/1612.06342}{{\ttfamily 1612.06342}}.

\bibitem{Cachazo:2016njl}
F.~Cachazo, P.~Cha and S.~Mizera, \emph{{Extensions of Theories from Soft
  Limits}}, \href{http://dx.doi.org/10.1007/JHEP06(2016)170}{\emph{JHEP}
  {\bfseries 06} (2016) 170},
  [\href{https://arxiv.org/abs/1604.03893}{{\ttfamily 1604.03893}}].

\bibitem{Kleiss:1988ne}
R.~Kleiss and H.~Kuijf, \emph{{Multi - Gluon Cross-sections and Five Jet
  Production at Hadron Colliders}},
  \href{http://dx.doi.org/10.1016/0550-3213(89)90574-9}{\emph{Nucl. Phys.}
  {\bfseries B312} (1989) 616--644}.

\bibitem{Nguyen:2009jk}
D.~Nguyen, M.~Spradlin, A.~Volovich and C.~Wen, \emph{{The Tree Formula for MHV
  Graviton Amplitudes}},
  \href{http://dx.doi.org/10.1007/JHEP07(2010)045}{\emph{JHEP} {\bfseries 07}
  (2010) 045}, [\href{https://arxiv.org/abs/0907.2276}{{\ttfamily 0907.2276}}].

\bibitem{Feng:2012sy}
B.~Feng and S.~He, \emph{{Graphs, determinants and gravity amplitudes}},
  \href{http://dx.doi.org/10.1007/JHEP10(2012)121}{\emph{JHEP} {\bfseries 10}
  (2012) 121}, [\href{https://arxiv.org/abs/1207.3220}{{\ttfamily 1207.3220}}].

\bibitem{Chiodaroli:2014xia}
M.~Chiodaroli, M.~Gunaydin, H.~Johansson and R.~Roiban, \emph{{Scattering
  amplitudes in $ \mathcal{N}=2 $ Maxwell-Einstein and Yang-Mills/Einstein
  supergravity}}, \href{http://dx.doi.org/10.1007/JHEP01(2015)081}{\emph{JHEP}
  {\bfseries 01} (2015) 081},
  [\href{https://arxiv.org/abs/1408.0764}{{\ttfamily 1408.0764}}].

\bibitem{Chiodaroli:2015rdg}
M.~Chiodaroli, M.~Gunaydin, H.~Johansson and R.~Roiban, \emph{{Spontaneously
  Broken Yang-Mills-Einstein Supergravities as Double Copies}},
  \href{https://arxiv.org/abs/1511.01740}{{\ttfamily 1511.01740}}.

\bibitem{Bern:2008qj}
Z.~Bern, J.~J.~M. Carrasco and H.~Johansson, \emph{New relations for
  gauge-theory amplitudes},
  \href{http://dx.doi.org/10.1103/PhysRevD.78.085011}{\emph{Phys. Rev. D}
  {\bfseries 78} (2008) 085011},
  [\href{https://arxiv.org/abs/0805.3993}{{\ttfamily 0805.3993}}].

\bibitem{Bern:2010ue}
Z.~Bern, J.~J.~M. Carrasco and H.~Johansson, \emph{{Perturbative Quantum
  Gravity as a Double Copy of Gauge Theory}},
  \href{http://dx.doi.org/10.1103/PhysRevLett.105.061602}{\emph{Phys. Rev.
  Lett.} {\bfseries 105} (2010) 061602},
  [\href{https://arxiv.org/abs/1004.0476}{{\ttfamily 1004.0476}}].

\bibitem{Cheung:2016prv}
C.~Cheung and C.-H. Shen, \emph{{Symmetry and Action for Flavor-Kinematics
  Duality}},  \href{https://arxiv.org/abs/1612.00868}{{\ttfamily 1612.00868}}.

\bibitem{Du:2016tbc}
Y.-J. Du and C.-H. Fu, \emph{{Explicit BCJ numerators of nonlinear simga
  model}}, \href{http://dx.doi.org/10.1007/JHEP09(2016)174}{\emph{JHEP}
  {\bfseries 09} (2016) 174},
  [\href{https://arxiv.org/abs/1606.05846}{{\ttfamily 1606.05846}}].

\bibitem{Carrasco:2016ldy}
J.~J.~M. Carrasco, C.~R. Mafra and O.~Schlotterer, \emph{{Abelian Z-theory:
  NLSM amplitudes and alpha'-corrections from the open string}},
  \href{https://arxiv.org/abs/1608.02569}{{\ttfamily 1608.02569}}.

\bibitem{Carrasco:2016ygv}
J.~J.~M. Carrasco, C.~R. Mafra and O.~Schlotterer, \emph{{Semi-abelian
  Z-theory: NLSM+phi{\textasciicircum}3 from the open string}},
  \href{https://arxiv.org/abs/1612.06446}{{\ttfamily 1612.06446}}.

\bibitem{Geyer:2015bja}
Y.~Geyer, L.~Mason, R.~Monteiro and P.~Tourkine, \emph{{Loop Integrands for
  Scattering Amplitudes from the Riemann Sphere}},
  \href{http://dx.doi.org/10.1103/PhysRevLett.115.121603}{\emph{Phys. Rev.
  Lett.} {\bfseries 115} (2015) 121603},
  [\href{https://arxiv.org/abs/1507.00321}{{\ttfamily 1507.00321}}].

\bibitem{Geyer:2015jch}
Y.~Geyer, L.~Mason, R.~Monteiro and P.~Tourkine, \emph{{One-loop amplitudes on
  the Riemann sphere}},
  \href{http://dx.doi.org/10.1007/JHEP03(2016)114}{\emph{JHEP} {\bfseries 03}
  (2016) 114}, [\href{https://arxiv.org/abs/1511.06315}{{\ttfamily
  1511.06315}}].

\bibitem{Cachazo:2015aol}
F.~Cachazo, S.~He and E.~Y. Yuan, \emph{{One-Loop Corrections from Higher
  Dimensional Tree Amplitudes}},
  \href{http://dx.doi.org/10.1007/JHEP08(2016)008}{\emph{JHEP} {\bfseries 08}
  (2016) 008}, [\href{https://arxiv.org/abs/1512.05001}{{\ttfamily
  1512.05001}}].

\bibitem{Chiodaroli:2017ngp}
M.~Chiodaroli, M.~Gunaydin, H.~Johansson and R.~Roiban, \emph{{Explicit
  Formulae for Yang-Mills-Einstein Amplitudes from the Double Copy}},
  \href{https://arxiv.org/abs/1703.00421}{{\ttfamily 1703.00421}}.

\end{thebibliography}\endgroup




\end{document}